\documentclass[a4paper,11pt]{article}
\usepackage{jheppub}
\usepackage[utf8]{inputenc}

\usepackage{color}
\definecolor{dark_red}{rgb}{.79, .39, .09}
\definecolor{pale_red}{rgb}{.92, .54, .25}
\definecolor{light_yellow}{rgb}{.95, .73, .57}

\definecolor{mypurple}{rgb}{0.49,0.18,0.56}
\definecolor{mygold}{rgb}{0.93,0.49,0.13}
\definecolor{mygreen}{rgb}{0,0.5,0}
\definecolor{myblue}{rgb}{0,0,0.75}
\definecolor{mymagenta}{cmyk}{0,1,0,0.12}
\definecolor{mygray}{rgb}{0.5,0.5,0.5}

\usepackage{graphicx}
\graphicspath{{tikz/}}
\usepackage[labelformat=simple]{subcaption}

\usepackage{tikz}
\usetikzlibrary{arrows}
\usetikzlibrary{positioning}
\usetikzlibrary{patterns}
\usetikzlibrary{calc}

\newcommand{\midarrow}{\tikz \draw[-triangle 90] (0,0) -- +(.1,0);}

\newsavebox{\measurebox}

\newcommand\tzunit{}
\newcommand\tzaux{}
\newcommand{\tzwidth}[1]{\def\tzaux{#1}}

\newenvironment{tzket}[1][]
{
    \def\tzunit{1}%
    \def\tzangle{75};
    \begin{scope}[#1]
        \draw[thick] ({-.5*\tzunit},{-.5*\tzunit}) -- ++(90:{2*\tzunit});
}{
        \draw[thick] ({\tzaux+.5*\tzunit},{-.5*\tzunit}) 
        -- ++(\tzangle:{\tzunit/sin(\tzangle)}) 
        -- ++({180-\tzangle}:{\tzunit/sin(\tzangle)});
    \end{scope}
}
\newenvironment{tzbraket}[1][]
{
    \def\tzunit{1}%
    \def\tzangle{75};
    \begin{scope}[#1]
        \draw[thick] ({0.5*\tzunit},{-.5*\tzunit}) 
        -- ++({180-\tzangle}:{\tzunit/sin(\tzangle)}) 
        -- ++(\tzangle:{\tzunit/sin(\tzangle)});
}{
        \draw[thick] ({\tzaux+.5*\tzunit},{-.5*\tzunit}) 
        -- ++(\tzangle:{\tzunit/sin(\tzangle)}) 
        -- ++({180-\tzangle}:{\tzunit/sin(\tzangle)});
    \end{scope}
}

\usepackage{amsfonts}
\usepackage{amssymb}
\usepackage{amsmath}
\usepackage{braket}
\usepackage{bbold}
\usepackage{bm}
\usepackage{nicefrac}

\newcommand{\lat}[1]{\mathcal{#1}}
\newcommand{\hilb}{\mathcal{H}}
\newcommand{\modes}{\mathcal{M}}
\newcommand{\falg}[1]{\lat{#1}_\pi}
\newcommand{\sss}[2][]{Z_{\lat{#2}#1}}
\newcommand{\sssi}[2]{Z_{\lat{#1}_#2}}
\newcommand{\hsss}[2][]{\hilb_\lat{#2}\!\left(\sss[#1]{#2}\right)}
\newcommand{\galg}[1]{\lat{#1}_g}
\newcommand{\plaqop}[1]{\hat{U}_{#1}}
\newcommand{\plaqob}[1]{\hat{\mathcal{U}}_{#1}}

\newcommand{\pathop}[1]{\hat{P}_{#1}}
\newcommand{\pathob}[1]{\hat{\mathcal{P}}_{#1}}

\usepackage{setspace}   % Allows to set the space between lines;
\usepackage{booktabs}   % allows to use \toprule, \midrule and \bottomrule in the tabular environment;
\usepackage{multirow}
\usepackage{umoline}
\usepackage{dcolumn}% Align table columns on decimal point

\begin{document}

\title{Entanglement Witnessing for Lattice Gauge Theories}

\author[a,1]{Veronica Panizza\note{Corresponding author.}}
\author[a,b,c]{Ricardo Costa de Almeida}
\author[a,c]{Philipp Hauke}

\affiliation[a]{INO-CNR BEC Center and Department of Physics, University of Trento, Via Sommarive 14, 38123 Trento, Italy}
\affiliation[b]{Institute for Theoretical Physics, Heidelberg University, Philosophenweg 16, 69120 Heidelberg, Germany}
\affiliation[c]{INFN-TIFPA, Trento Institute for Fundamental Physics and Applications, Trento, Italy}

\emailAdd{veronica.panizza@unitn.it}
\emailAdd{r.costadealmeida@unitn.it}
\emailAdd{philipp.hauke@unitn.it}

\date{\today}

\abstract{
    Entanglement is assuming a central role in modern quantum many-body physics.
    Yet, for lattice gauge theories its certification remains extremely challenging.
    A key difficulty stems from the local gauge constraints underlying the gauge theory, which separate the full Hilbert space into a direct sum of subspaces characterized by different superselection rules. 
    In this work, we develop the theoretical framework of entanglement witnessing for lattice gauge theories that takes this subtlety into account. 
    We illustrate the concept at the example of a \(\mathrm{U}(1)\) lattice gauge theory in 2+1 dimensions, without and with dynamical fermionic matter.
    As this framework circumvents costly state tomography, it opens the door to resource-efficient certification of entanglement in theoretical studies as well as in laboratory quantum simulations of gauge theories. 
}

\maketitle

\section{Introduction}
\label{sec:intro}

In addition to its large conceptual importance for the foundations of quantum theory, entanglement also constitutes a key resource for quantum technologies \cite{friis_vitagliano_malik_huber_2018,pezze_smerzi_oberthaler_schmied_treutlein_2018,acn_bloch_buhrman_calarco_eichler_eisert_esteve_gisin_glaser_jelezko,azzini_mazzucchi_moretti_pastorello_pavesi_2020} and plays an important role in strongly-correlated quantum many-body systems \cite{nandkishore_huse_2015,alet_laflorencie_2018,abanin_altman_bloch_serbyn_2019,wen_2019}.
Historically, entanglement has been studied widely in systems consisting of spins (i.e., qubits or qudits) or similar local degrees of freedom that interact with each other \cite{amico_fazio_osterloh_vedral_2008,horodecki_horodecki_horodecki_horodecki_2009,laflorencie_2016}.
The characterization of entanglement is, however, much more subtle in highly-constrained systems such as lattice gauge theories (LGTs) \cite{buividovich_polikarpov_2008,donnelly_2012,casini_huerta_rosabal_2014,gromov_santos_2014,ghosh_soni_trivedi_2015,Soni2016JHEP,aoki_iritani_nozaki_numasawa_shiba_tasaki_2015,radicevic_2016,van_acoleyen_bultinck_haegeman_marien_scholz_verstraete_2016,radicevic_2022,abrahamsen_su_tong_wiebe_2022}.
Such theories are defined by local conservation laws given by the gauge symmetry (e.g., the Gauss's law in quantum electrodynamics (QED)).
These restrict the dynamics to small parts of the full Hilbert space, so-called superselection sectors (SSs), which are determined by an extensive set of local conserved charges \cite{naaijkens2014superselection,casini_huerta_magan_pontello_2020}.
As a consequence, coherences between two SSs with different local charges cannot be accessed, as there is no physical, gauge-invariant operation that would allow one to couple them.
This generates a situation that is drastically different from spin systems, where arbitrary operations are allowed to explore the full Hilbert space.

Although challenging, it is highly desirable to get a handle on entanglement in LGTs: 
on the theoretical side, the role of entanglement in the thermalization and equilibration of gauge theories is an outstanding question in the field of high-energy physics \cite{berges_floerchinger_venugopalan_2018,bellwied_2018,feal_pajares_vazquez_2019,Mueller2022PRL}; 
on the practical side, gauge theories are of great interest for quantum technologies, e.g., through their close relation to paradigms of topological quantum computing \cite{lechner_hauke_zoller_2015} and the large number of gauge quantum-simulation experiments that are currently being developed \cite{martinez_muschik_schindler_nigg_erhard_heyl_hauke_dalmonte_monz_zoller_et_al_2016,bernien_schwartz_keesling_levine_omran_pichler_choi_zibrov_endres_greiner_et_al_2017,klco_dumitrescu_mccaskey_morris_pooser_sanz_solano_lougovski_savage_2018,yang_sun_ott_wang_zache_halimeh_yuan_hauke_pan_2020,banuls_blatt_catani_celi_cirac_dalmonte_fallani_jansen_lewenstein_montangero_et_al_2020,mil_zache_hegde_xia_bhatt_oberthaler_hauke_berges_jendrzejewski_2020,alexeev_bacon_brown_calderbank_carr_chong_demarco_englund_farhi_fefferman_et_al_2021,aidelsburger_barbiero_bermudez_chanda_dauphin_et_al_2021,zohar_2021,mildenberger_mruczkiewicz_halimeh_jiang_hauke_2022,nguyen_tran_zhu_green_alderete_davoudi_linke_2022,wang_khatami_fei_wyrick_namboodiri_kashid_rigosi_bryant_silver_2022}.
In recent years, the subtleties of defining entanglement entropy and other full-fledged entanglement measures for LGTs have been highlighted in the literature \cite{buividovich_polikarpov_2008,donnelly_2012,gromov_santos_2014,casini_huerta_rosabal_2014,aoki_iritani_nozaki_numasawa_shiba_tasaki_2015,ghosh_soni_trivedi_2015,Soni2016JHEP,van_acoleyen_bultinck_haegeman_marien_scholz_verstraete_2016,radicevic_2016,radicevic_2022,abrahamsen_su_tong_wiebe_2022}.
However, the experimental observation of such entanglement measures requires knowledge of the system state that grows exponentially with system size.
In contrast, the concept of entanglement witnessing \cite{Horodecki1996PLA,Terhal2000PLA,guhne_toth_2009,chruscinski_sarbicki_2014,baker_kharzeev_2018,lewenstein_kraus_cirac_horodecki_2000} provides an efficient and scalable framework for accessing entanglement.
However, it has not yet been extended to LGTs.

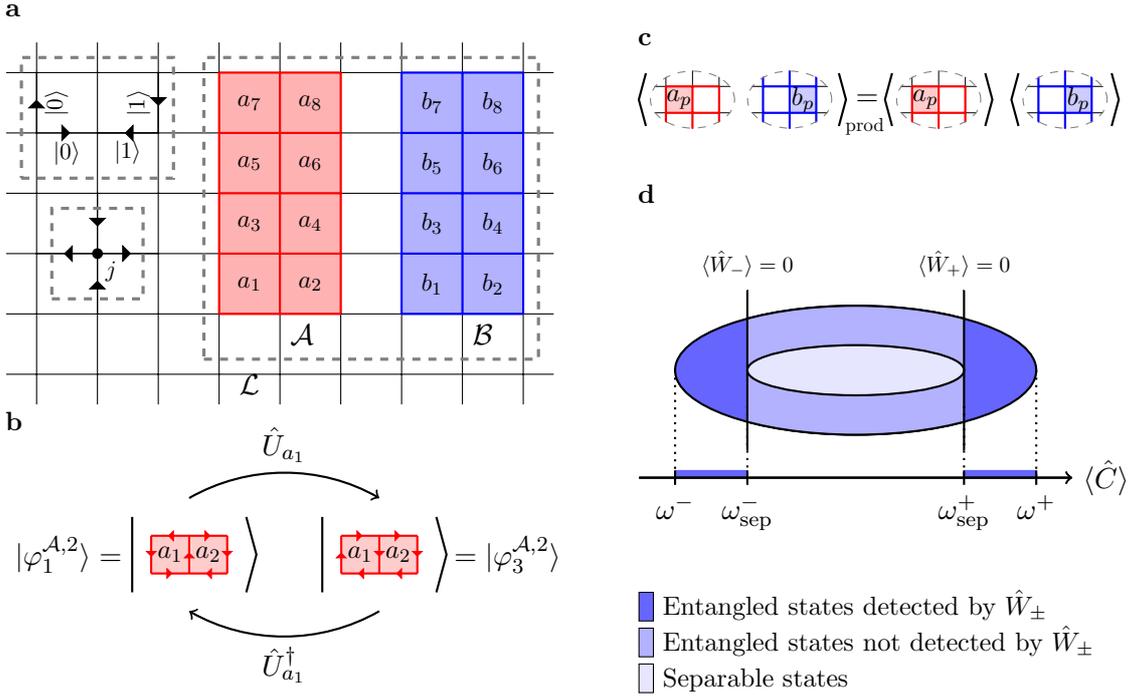
\begin{figure}[t]
    \centering
    \begin{minipage}{0.5\textwidth}
        \begin{subfigure}[t]{\textwidth}
            \caption{}
            \begin{tikzpicture}[scale=.8]
    \draw[fill,color=red!30] (0,0) rectangle (2,4);
    \draw[fill,color=blue!30] (3,0) rectangle (5,4);
    \draw[thin] (-3.5,-1.5) grid (5.5,4.5);
    \draw[red, thick] (0,0) grid (2,4);
    \draw[blue, thick] (3,0) grid (5,4);
    \draw (1,0) node[anchor=north west]{$\mathcal{A}$};
    \draw (4,0) node[anchor=north west]{$\mathcal{B}$};
    \draw (0.5,-1.5) node[anchor=south]{$\mathcal{L}$};
    \begin{scope}[every node/.style={scale=.85}]
        \draw (.5,.5) node{$a_1$};
        \draw (1.5,.5) node{$a_2$};
        \draw (.5,1.5) node{$a_3$};
        \draw (1.5,1.5) node{$a_4$};
        \draw (.5,2.5) node{$a_5$};
        \draw (1.5,2.5) node{$a_6$};
        \draw (.5,3.5) node{$a_7$};
        \draw (1.5,3.5) node{$a_8$};
        
        \draw (3.5,.5) node{$b_1$};
        \draw (4.5,.5) node{$b_2$};
        \draw (3.5,1.5) node{$b_3$};
        \draw (4.5,1.5) node{$b_4$};
        \draw (3.5,2.5) node{$b_5$};
        \draw (4.5,2.5) node{$b_6$};
        \draw (3.5,3.5) node{$b_7$};
        \draw (4.5,3.5) node{$b_8$};
    \end{scope}
    \begin{scope}[every node/.style={allow upside down, sloped,scale=.8}]
        \draw (-3,3) -- node{\midarrow} (-3,4) node[pos=.5, anchor=north]{$\ket{0}$};
        \draw (-1,4) -- node{\midarrow} (-1,3);
        \draw[very thin] (-1,3) -- (-1,4) node[pos=.5,anchor=south]{$\ket{1}$};
        \draw (-3,3) -- node{\midarrow} (-2,3) node[pos=.5,anchor=north]{$\ket{0}$};
        \draw (-1,3) -- node{\midarrow} (-2,3);
        \draw[very thin] (-2,3) -- node[anchor=north]{$\ket{1}$} (-1,3);
        
        \draw (-2,0) -- node{\midarrow} (-2,1);
        \draw (-2,1) -- node{\midarrow} (-1,1);
        \draw (-2,2) -- node{\midarrow} (-2,1);
        \draw (-2,1) -- node{\midarrow} (-3,1);
    \end{scope}
    \draw[dashed,gray,very thick] (-3.25,2.25) rectangle (-0.75,+4.25);
    \draw[dashed,gray,very thick] (-2.75,1.75) rectangle (-1.25,0.25);
    \draw[dashed,gray,very thick] (-.25,-.75) rectangle (5.25,4.25);
    \draw[fill] (-2,1)  circle (.08) node[anchor=north west,scale=.8]{$j$};
\end{tikzpicture}
            \label{fg:main:lattice}
        \end{subfigure}
        \begin{subfigure}[t]{\textwidth}
            \caption{}
            \vspace{-5mm}
            \begin{tikzpicture}[scale=.5]
    \draw (-.5,.5) node[anchor=east,scale=1]{$\ket{\varphi^{\mathcal{A},2}_1} =$};
    \begin{tzket}
        \draw[fill,color=red!20] (0,0) rectangle (2,1);
        \draw (.5,.5) node[scale=.9]{$a_1$};
        \draw (1.5,.5) node[scale=.9]{$a_2$};
        \begin{scope}[thick,red,every node/.style={allow upside down, sloped, scale=.5}]
            \draw (0,0) -- node{\midarrow} (1,0);
            \draw (1,0) -- node{\midarrow} (1,1);
            \draw (1,1) -- node{\midarrow} (0,1);
            \draw (0,1) -- node{\midarrow} (0,0);
            \draw (1,1) -- node{\midarrow} (2,1);
            \draw (2,1) -- node{\midarrow} (2,0);
            \draw (2,0) -- node{\midarrow} (1,0);
        \end{scope}
        \tzwidth{2}
    \end{tzket}
    \draw[thick,->] (1,2) arc(120:60:5) node[pos=.5,anchor=south]{$\hat{U}_{a_1}$};
    \draw[thick,<-] (1,-1) arc(240:300:5) node[pos=.5,anchor=north]{$\hat{U}_{a_1}^\dag$};
    
    \def \S {5};
    \begin{tzket}[shift={(5,0)}]
        \draw[fill,color=red!20] (0,0) rectangle (2,1);
        \draw (+.5,.5) node[scale=.9]{$a_1$};
        \draw (+1.5,.5) node[scale=.9]{$a_2$};
        \begin{scope}[thick, red,every node/.style={allow upside down, sloped, scale=.5}]
            \draw (1,0) -- node{\midarrow} (0,0);
            \draw (1,1) -- node{\midarrow} (1,0);
            \draw (0,1) -- node{\midarrow} (1,1);
            \draw (0,0) -- node{\midarrow} (0,1);
            \draw (1,1) -- node{\midarrow} (2,1);
            \draw (2,1) -- node{\midarrow} (2,0);
            \draw (2,0) -- node{\midarrow} (1,0);
        \end{scope}
        \tzwidth{2}
    \end{tzket}
    \draw (7.7,.5) node[anchor=west,scale=1]{$= \ket{\varphi^{\mathcal{A},2}_3}$};
\end{tikzpicture}
            \label{fg:main:plaquette}
        \end{subfigure}
    \end{minipage}
    \hfill
    \begin{minipage}{0.45\textwidth}
        \begin{subfigure}[t]{\textwidth}
            \caption{}
            \begin{tikzpicture}[scale=.35]
    \begin{tzbraket}[shift={(-.5,0)}]
        \begin{scope}[shift={(2.25,0)}]
            \clip (0,.5) ellipse (1.6 and 1.1);
            \draw[very thin] (-2.5,-2.5) grid (4,4);
            \draw[fill,red!20] (-1,0) rectangle (0,1);
            \draw (-.5,.5) node[scale=.9]{$a_p$};
            \draw[thick,red] (-1,-2) grid (1,1);
            \draw[gray,dashed,thick]  (0,.5) ellipse (1.6 and 1.1);
        \end{scope}
        \def \s{3.4};
        \begin{scope}[shift={(2.5,0)}]
            \clip (\s,.5) ellipse (1.6 and 1.1);
            \draw[xshift=-.6cm] (\s-2.5,-2.5) grid +(\s+4,4);
            \draw[fill,blue!20] (\s,0) rectangle (\s+1,1);
            \draw (\s+.5,.5) node[scale=.9]{$b_p$};
            \draw[thick,blue,xshift=-.6cm] (3,-1) grid +(2,3);
            \draw[gray,dashed,thick]  (\s,.5) ellipse (1.6 and 1.1);
        \end{scope}
        \tzwidth{7.25}
    \end{tzbraket}
    \draw (7.25,-.5) node[anchor=west,scale=0.65]{$\mathrm{prod}$};
    \draw (7.5,.5) node[anchor=west,scale=1]{$=$};
    \begin{tzbraket}[shift={(8.75,0)}]
        \begin{scope}[shift={(2.25,0)}]
            \clip (0,.5) ellipse (1.6 and 1.1);
            \draw[very thin] (-2.5,-2.5) grid (4,4);
            \draw[fill,red!20] (-1,0) rectangle (0,1);
            \draw (-.5,.5) node[scale=.9]{$a_p$};
            \draw[thick,red] (-1,-2) grid (1,1);
            \draw[gray,dashed,thick] (0,.5) ellipse (1.6 and 1.1);
        \end{scope}
         \tzwidth{3.5}
    \end{tzbraket}
    \begin{tzbraket}[shift={(13.5,0)}]
        \begin{scope}[shift={(2.25,0)}]
            \clip (0,.5) ellipse (1.6 and 1.1);
            \draw (-2.5,-2.5) grid (4,4);
            \draw[fill,blue!20] (0,0) rectangle (1,1);
            \draw (0.5,.5) node[scale=.9]{$b_p$};
            \draw[thick,blue] (-1,-1) grid +(2,3);
            \draw[gray,dashed,thick] (0,.5) ellipse (1.6 and 1.1);
        \end{scope}
         \tzwidth{3.5}
    \end{tzbraket}
    \end{tikzpicture}
            \label{fg:main:separable}
            \vspace{5mm}
        \end{subfigure}
        \begin{subfigure}[t]{\textwidth}
            \caption{}
            \vspace{-5mm}
            \begin{tikzpicture}[scale=.95] 
    \filldraw[draw=black,thick,fill=blue!60] (4,1.5) ellipse (2.5 and .9);
    \begin{scope}
        \clip (2.5,0) rectangle (5.5,4);
        \filldraw[draw=black,thick,fill=blue!30] (4,1.5) ellipse (2.5 and .9);
    \end{scope}
    
    \filldraw[draw=black,thick,fill=blue!10] (4,1.5) ellipse(1.5 and .35);
    
    \begin{scope}
        \clip (4,1.5) ellipse (2.5 and 1.4);
        \draw[thick] (2.5,3) -- (2.5,0);
        \draw[thick] (5.5,3) -- (5.5,0);
    \end{scope}
    \draw (2.5,2.7) node[anchor=south,scale=.75]{$\langle\hat{W}_-\rangle = 0$};
    \draw (5.5,2.7) node[anchor=south,scale=.75]{$\langle\hat{W}_+\rangle = 0$};
    
    \draw[thick,dotted] (2.5,1) -- (2.5,0);
    \draw[thick,dotted] (5.5,1) -- (5.5,0);
    \draw[thick,dotted] (1.5,1.5) -- (1.5,0);
    \draw[thick,dotted] (6.5,1.5) -- (6.5,0);
    
    \draw[fill,color=blue!60] (1.5,.1) rectangle (2.5,0);
    \draw[fill,color=blue!60] (5.5,.1) rectangle (6.5,0);
    
    \draw[thick,->] (1,0) -- (7,0) node[anchor=west]{$\langle\hat{C}\rangle$};
     \draw[thick,->] (1,0) -- (7,0);
    \draw[thick] (1.5,.1) -- (1.5,-.1) node[anchor=north]{$\omega^-$};
    \draw[thick] (6.5,.1) -- (6.5,-.1) node[anchor=north]{$\omega^+$};
    
    \draw[thick] (2.5,.1) -- (2.5,-.1) node[anchor=north]{$\omega_\mathrm{sep}^-$};
    \draw[thick] (5.5,.1) -- (5.5,-.1) node[anchor=north]{$\omega_\mathrm{sep}^+$};
    
    \filldraw[draw = black, fill = blue!60] (1,-2) rectangle (1.2,-1.6);
    \draw (1.2,-1.78) node[anchor=west,scale=.9]{Entangled states detected by $\hat{W}_\pm$}; 
    
    \filldraw[draw=black,fill=blue!30] (1,-2.5) rectangle(1.2,-2.1);
    \draw (1.2,-2.28) node[anchor=west,scale=.9]{Entangled states not detected by $\hat{W}_\pm$};
    
    \filldraw[draw=black,fill=blue!10] (1,-3) rectangle (1.2,-2.6);
    \draw (1.2,-2.82) node[anchor=west,scale=.9]{Separable states};
\end{tikzpicture}
            \label{fg:main:witness}
        \end{subfigure}
    \end{minipage}
    \caption{
        Entanglement witnessing in lattice gauge theories.
        \textbf{a)} Lattice used for the examples in this work.
        Top left: Basis choice for the two-dimensional Hilbert space associated with the electric field living on edges.
        Bottom left: Physical states obey a local gauge symmetry, here exemplified for a \(\mathrm{U}(1)\) pure gauge theory.
        At all vertices \(j\), adjacent electric fields have to sum up to zero (\textit{two-in-two-out rule}).
        Right: Example of disjoint subregions \(\lat{A}\) and \(\lat{B}\) for which entanglement witnesses are constructed in this work (\(a_p\) and \(b_p\) label plaquettes).
        \textbf{b)} The expectation value of operator \(\plaqop{a_1}\) (\(\plaqop{a_1}^\dag\)) on \(\ket{\Psi_\lat{A}}\) as defined in equation~\eqref{eq:gauge:decomposition} is \(\alpha_1\alpha_3^*\) (\(\alpha_1^* \alpha_3\)).
        Analogous calculations apply to each operator in equation~\eqref{eq:gauge:2pwitness}.
        \textbf{c)} We define separable states as those where all products of operators \(\hat{O}_\lat{A}\) and \(\hat{O}_\lat{B}\), acting on \(\lat{A}\) and \(\lat{B}\), respectively, factorize.
        Witness operators \(\hat{W}_\pm\) are constructed from sums of such products, \(\hat{C}=\sum_m\hat{O}^m_\lat{A}\hat{O}^m_\lat{B}\) (see equation~\eqref{eq:witness:operator}).
        \textbf{d)} If constructed suitably, bounds for the expectation value of $\hat{C}$ on generic states exceed those of separable states and equation~\eqref{eq:witness:operator} defines effective entanglement witnesses $\hat{W}_\pm$.
    }
    \label{fg:main}
\end{figure}

In this work, we introduce entanglement witnessing for LGTs.
This framework is based on a careful definition of separability of such a highly constrained system.
We obtain a viable definition by combining recent developments concerning entanglement in LGTs with a physically motivated approach that has, e.g., been successfully applied in the past to massive fermions that are subject to the \emph{global} superselection constraint of particle number conservation \cite{banuls_cirac_wolf_2007,zanardi_2002,bourennane_eibl_kurtsiefer_gaertner_weinfurter_guhne_hyllus_brus_lewenstein_sanpera_2004}.
In our framework, two spatial regions of a LGT (figures~\ref{fg:main:lattice}) are separable if and only if the expectation values of all gauge invariant observables with support in both regions decompose into products (see figure~\ref{fg:main:separable}).
Conversely, if the expectation value of any such observable exceeds the expectation value allowed by separable regions, the state of the system is witnessed as entangled (see figure~\ref{fg:main:witness}).
Using the example of a \(\mathrm{U}(1)\) LGT in two spatial dimensions, we analytically illustrate the role of the SSs, as these crucially determine the attainable expectation values, and we provide an example for an entanglement witness in this LGT.
In a second example, we illustrate the entanglement witnessing framework for a LGT consisting of \(\mathrm{U}(1)\) gauge fields coupled to dynamical fermionic particles and anti-particles.
Our discussions serve to showcase how some of the subtleties in detecting entanglement in LGTs can be overcome with limited practical resources.
 
The rest of this paper organized as follows: In section~\ref{sec:formal}, we discuss an operative approach to define separable states in situations where the system is subject to a symmetry constraint.
This definition permits us to define adequate entanglement witnesses.
We illustrate this concept at the examples of a fermionic system with parity conservation and a general LGT with arbitrary gauge symmetry.
In section~\ref{sec:gauge}, we specialize to the example of a pure \(\mathrm{U}(1)\) lattice gauge theory in 2+1D.
We introduce a set of entanglement witnesses, exemplify its effectiveness through an explicit example of a small system, and discuss the scaling of required resources with system size.
In section~\ref{sec:matter}, we extend the scenario of section~\ref{sec:gauge} by coupling the \(\mathrm{U}(1)\) gauge field to fermionic matter.
Finally, section~\ref{sec:discussion} presents our conclusions and discussion of potential lines of inquiry.
The paper is complemented by Appendices that contain further technical details on conditions for a SS to be physical, a scheme that enables to remove frozen plaquettes from the considerations, and the numerical optimization of expectation-value bounds of the considered witnesses.

\section{Separability and Entanglement in Lattice Gauge Theories}  
\label{sec:formal}

Entanglement is a fundamental feature of composite quantum systems.
It relates to the amount of information and correlations shared in a non--classical manner between different constituents \cite{horodecki_horodecki_horodecki_horodecki_2009}.

For a system composed of local \(d\)-dimensional degrees of freedom (qudits), it is straightforward to define entanglement, since the many--body Hilbert space has a decomposition into a tensor product.
In particular, one can easily identify product states, i.e., pure states that do not have any quantum correlations.
More concretely, product states are those states that can be written as a tensor product,
\begin{equation}
    \ket{\psi_\text{prod}}
    = \bigotimes_{j\in \lat{L}} \ket{\psi_j}
    \in\hilb 
    = \bigotimes_{j\in \lat{L}} \hilb_j 
    \,\text{,} 
    \label{eq:qudit:product}
\end{equation}
where each \(\ket{\psi_j}\in\hilb_j\simeq\mathbb{C}^d\) denotes the local state with respect to some labeling \(\lat{L}\) of the physical qudits.
Separable states constitute a more general set of non--entangled states which includes mixed states with classical correlations.
Nonetheless, they are simply characterized by product states since all separable state \(\rho_\mathrm{sep}\) can be written as a convex combination,
\begin{equation}
    \rho_\text{sep}
    =\sum_\mu p_\mu \ket{\mu_\text{prod}}\bra{\mu_\text{prod}}
    \,\text{,} 
    \label{eq:qudit:separable}
\end{equation}
of product states \(\ket{\mu_\text{prod}}\).
Crucially, any state not of this form is entangled, so we can leverage information about product states to detect entangled states.

Regardless of the underlying degrees of freedom (i.e., the dimension of the \(\hilb_j\) and associated algebra), whenever a tensor product decomposition exists, one can use equation~\eqref{eq:qudit:product} to pinpoint non--entangled states.
However, there are physical systems that do not admit such a decomposition, as is the case for the LGTs considered in this paper.
For such systems, the definition of product/separable states requires some care.
Specifically, one has to use the factorization of expectation values that holds for all product state,
\begin{equation}
    \bra{\psi_\mathrm{prod}} \prod_{j\in \lat{L}} \hat{O}_j \ket{\psi_\mathrm{prod}}
    = \prod_{j\in \lat{L}} \braket{\psi_j|\hat{O}_j |\psi_j}
    \,\text{,}
    \label{eq:qudit:factorize}
\end{equation}
as the basis for distinguishing non-entangled and entangled states.

\subsection{Fermions with particle--number conservation}
\label{sec:formal:fermion}

Before tackling the case of LGTs, it is instructive to consider the illustrative and pedagogical example of a fermionic many--body system.
In particular, it provides a familiar example of a system whose product states are not described by equation~\eqref{eq:qudit:product}.
This will shed light in the subtleties that occur in presence of superselection rules and the implications for the separability criteria.

The Hilbert space of a system of fermionic particles is a Fock space,
\begin{equation}
    \hilb =\bigoplus_{N=0}^\infty \hilb_N 
    =\bigoplus_{N=0}^\infty 
    \underbrace{
        \hilb_{\mathrm{sp}}\wedge\dots\wedge\hilb_{\mathrm{sp}}
    }_{N\text{th exterior power}}
    \,\text{,} 
    \label{eq:fermion:hilbert}
\end{equation}
generated by the exterior powers of the single--particle Hilbert space \(\mathcal{H}_{\mathrm{sp}}\).
Due to the antisymmetric nature of the exterior product, it does not admit a tensor product decomposition that is compatible with the fermionic operator algebra.
This is necessary to account for the canonical anticommutation relations, but it is an obstacle as far as equation~\eqref{eq:qudit:product} is concerned.
Therefore, we require an alternative approach to define product states and identify entangled states.

As pointed out in reference~\cite{banuls_cirac_wolf_2007}, the solution comes from equation~\eqref{eq:qudit:factorize} as it is a criteria that relies solely on understanding the operator algebra and some well-defined notion of subsystems.
For instance, this approach provides and elegant way of understanding the notion of a partial trace for systems of indistinguishable particles \cite{balachandran_govindarajan_de_queiroz_reyes_lega_2013}.
Most importantly, to obtain a consistent rule it is critical to account for the parity superselection rules~\cite{banuls_cirac_wolf_2007,Friis2013PRA,Friis2016NJP}, as parity conservation informs how one must enforce equation~\eqref{eq:qudit:factorize}.
Concretely, given a bipartition of the system into subsystems \(\lat{A}\) and \(\lat{B}\), a product state is any state that fulfills  
\begin{equation}
    \braket{\hat{O}_\lat{A}\hat{O}_\lat{B}}_ \mathrm{prod}
    =\braket{\hat{O}_\lat{A}}_\lat{A}\braket{\hat{O}_\lat{B}}_\lat{B}
    \label{eq:fermion:factorize}
\end{equation}
for all physical observables of the form \(\hat{O}=\hat{O}_\lat{A}\hat{O}_\lat{B}\), where \(\hat{O}_\lat{A}\) only acts on \(\lat{A}\) and \(\hat{O}_\lat{B}\) only acts on \(\lat{B}\).
The expectation values in the right--hand side are with respect to local states.

In order to make sense of equation~\eqref{eq:fermion:factorize}, we must clarify what are subsystems and local observables in this context.
To do so, we fix a basis for \(\hilb_\mathrm{sp}\) so that there are modes \(\modes\) which label the single--particle basis states and the associated creation (annihilation) operators \(\hat{c}^\dagger_m\text{(}\hat{c}_m\text{)}\).
The bipartition of the system splits \(\modes\) into a disjoint union \(\modes=\modes_\lat{A}\sqcup \modes_\lat{B}\) with \(\modes_\lat{A}\) and \(\modes_\lat{B}\) defining the subsystems.
There are two algebras of operators, \(\falg{A}\) and \(\falg{B}\), associated to the bipartition that correspond to the local physical operators of \(\lat{A}\) and \(\lat{B}\), respectively.
Here, local means that an operator \(\hat{O}_\lat{A}\in\falg{A}\) can only affect \(\modes_\lat{A}\) modes, and similarly for operators in \(\falg{B}\).
As in~\cite{banuls_cirac_wolf_2007}, the \(\pi\) subscript indicates that the operators in \(\falg{A}\) and \(\falg{B}\) must also be compatible with the parity operator \(\hat{P}=(-1)^{\sum_m \hat{c}_m^\dagger\hat{c}_m}\) associated to the corresponding subsystem.
Hence, given a product state, equation~\eqref{eq:fermion:factorize} must be valid for all \(\hat{O}_\lat{A}\in\falg{A}\) and \(\hat{O}_\lat{B}\in\falg{B}\).
However, it does not need to hold for nonphysical observables that violate the parity superselection rule.

From equation~\eqref{eq:fermion:factorize}, one can define separable states of a fermionic many--body systems by considering convex combinations of product states, as in the case of a tensor-product structure of Hilbert space.
Again, it is important to enforce consistency with the parity superselection rule~\cite{banuls_cirac_wolf_2007,Friis2016NJP}.
As such, all states envolved must be physical, i.e., there can be no coherence between different superselection sectors of \(\hat{P}\).

\subsection{Separable states in LGTs}
\label{sec:formal:separable}

As the fermionic case illustrates, the algebras of local operators are the fundamental objects that inform the restrictions placed on product states.
In particular, the fewer observables qualify as physical, the weaker the constraint given by equation~\eqref{eq:fermion:factorize} becomes, thus making a larger set of states separable.
Hence, in a highly constraint system, fewer states qualify as entangled.
With this insight, we are ready to discuss the notion of separability in LGTs, which will present the theoretical basis for constructing entanglement witnesses for such systems.

Crucially, our goal is to describe entanglement as an observer who lives within the LGT would perceive it.
Hence, the gauge symmetry constitutes a fundamental, and inescapable, aspect of the system, and only gauge--invariant observables are admissible.
This is in sharp contrast with physical realizations that may exist, e.g., in an actual laboratory device that performs a gauge--theory quantum simulation.
An outside experimenter may perform certain rotations on the qubits constituting the device or perform measurements in basis that are incompatible with Gauss's law, which observers within the gauge theory have no access to.
However, for our purposes, any coherence that couples different SSs should be disregarded from the point of view of the gauge theory, since it does not translate into non--trivial quantum correlations.

The aim here is to detect entanglement between two subregions \(\lat{A}\) and \(\lat{B}\) of a LGT defined on a lattice \(\lat{L}\) (see figure~\ref{fg:main:lattice}).
For simplicity, \(\lat{A}\) and \(\lat{B}\) are taken to be two disjoint subregions whose boundaries do not touch directly.
This avoids subtleties that occur when dividing a LGT into subsystems whose boundaries do touch \cite{casini_huerta_rosabal_2014,casini_huerta_magan_pontello_2020,ghosh_soni_trivedi_2015,Soni2016JHEP} and simplifies some discussions and computations later on.
Nonetheless, our approach is generalizable.
Similarly to the fermionic case, there are algebras \(\galg{A}\) and \(\galg{B}\) that correspond to operators with support in \(\lat{A}\) and \(\lat{B}\).
The subscript \(g\) indicates that the algebra only includes operators that respect the underlying symmetry, meaning here that they are gauge invariant.
Just as before, product states are those that factorize according to equation~\eqref{eq:fermion:factorize} for all \(\hat{O}_\lat{A}\in \galg{A}\) and \(\hat{O}_\lat{B}\in \galg{B}\) (see figure~\ref{fg:main:separable}).
As one might expect, only gauge invariant states are allowed, in the full lattice and also in the subregions, so that the local expectation values are taken with respect to physical states.

In order to further characterize product states, we notice that \(\galg{A}\) and \(\galg{B}\) can have a non--trivial center~\cite{casini_huerta_rosabal_2014}.
In fact, this is precisely the obstruction to the existence of tensor product decomposition and we can use this to our advantage.
To make this concrete, let us consider a set of generators \(\{\hat{Z}_j\}_j\) of the center of \(\galg{A}\), denoted \(Z(\galg{A})\).
Since the generators can be diagonalized simultaneously, there is an eigenspace \(\hsss{A}\) for each set of eigenvalues \(\sss{A}=\{Z_j\}_j\), i.e., for all \(\ket{\psi_\lat{A}}\in\hsss{A}\) we have \(\hat{Z}_j\ket{\psi_\lat{A}}=Z_j\ket{\psi_\lat{A}}\).
The operators of \(\galg{A}\) do not couple different \(\hsss{A}\) since they commute with the elements of the center, so the eigenvalues \(\sss{A}\) amount to SSs of the subregion \(\lat{A}\).
Naturally, the same principle applies to \(\lat{B}\) and to the region covering the rest of the lattice, \(\lat{R}=\overline{\left(\lat{A}\cup\lat{B}\right)}\).
As a consequence, one has a decomposition of the Hilbert space of the LGT,
\begin{equation}
    \hilb
    =\bigoplus_{(\sss{A},\sss{R},\sss{B})} \hsss{A}\otimes \hsss{R} \otimes \hsss{B}
    \,\mathrm{,}
    \label{eq:gauge:hilbert}
\end{equation}
into a direct sum of the SSs \(\sss{A},\sss{R}\) and \(\sss{B}\).
Equation~\eqref{eq:gauge:hilbert} holds because inside the eigenspaces the generators of the centers are proportional to the identity, so that they act trivially and there is no obstruction to a tensor decomposition within a fixed SS.
However, there is an important caveat, namely the direct sum only runs over compatible combinations \((\sss{A},\sss{R},\sss{B})\) of superselection sectors.
For instance, the shared boundary of \(\lat{A}\) and \(\lat{R}\) implies that the generators of \(Z(\galg{A})\) are related to operators in \(Z(\galg{R})\) so that the values of \(\sss{A}\) and \(\sss{R}\) are not independent (and mutatis mutandis for  \(\sss{B}\) and \(\sss{R}\)).
Hence, the compatibility rules between \(\sss{A},\sss{R}\), and \(\sss{B}\) capture how, and when, one can glue gauge invariant states of the subregions to construct a global gauge invariant state.
In Section~\ref{sec:gauge}, where we discuss the concrete example of an \(\mathrm{U}(1)\) LGT, we give an explicit construction of the generators of the centers, and it will be clear that the superselection sectors cannot be combined arbitrarily.

Equation~\eqref{eq:gauge:hilbert} provides a way to use information about the algebras of locally gauge invariant operators to characterize separable states.
Namely, we know that the product states factor after the projection into the eigenspaces and can use this to calculate expectation values of  separable states.
Hence, once can write a separable mixed state restricted to \(\lat{A}\) and \(\lat{B}\) as
\begin{equation}
    \rho_{\mathrm{sep}}|_{\lat{A}\cup\lat{B}}
    = \bigoplus_{(\sss{A},\sss{B})} p(\sss{A},\sss{B})\rho_{\lat{A}}(\sss{A})\otimes \rho_\lat{B}(\sss{B})
    \,\mathrm{,}
    \label{eq:gauge:separable}
\end{equation}
where the direct sum runs over compatible SSs (compare to equation (26) in~\cite{casini_huerta_rosabal_2014}).
The distribution of the state into SSs is inherited from the global state and encoded into the probabilities \(p(Z_\mathcal{A},Z_\mathcal{B})\).
States of this form are exactly those that fulfil the requirement that correlators between physically allowed observables factorize as in  equation~\eqref{eq:fermion:factorize}.
In some sense, this turns the approach typical of tensor-product Hilbert spaces around, where first separability is defined on the state level from which then consequences on observables follow.
Importantly, equation~\eqref{eq:gauge:separable} does not imply that the overall Hilbert space is a tensor product as the non--trivial structure of equation~\eqref{eq:gauge:hilbert} manifests itself through the consistency relations among the different SSs.

Notice our discussion has no explicit mention of the superselection sectors of \(\lat{L}\) so far.
In principle, this can be accounted for by \(\sss{R}\).
Nonetheless, it is more convenient to fix the global superselection sector from the start, so that \(Z(\galg{L})\) becomes trivial.
Hence, unless otherwise mentioned, we work with a fixed superselection sector \(\hsss{L}\) instead of the full Hilbert space.

\subsection{Entanglement witnesses for LGTs}
\label{sec:formal:witness}

An entanglement witness~\cite{Horodecki1996PLA,Terhal2000PLA,guhne_toth_2009,chruscinski_sarbicki_2014,baker_kharzeev_2018,lewenstein_kraus_cirac_horodecki_2000} is an operator \(\hat{W}\) chosen such that the hyperplane \(\braket{\hat{W}}_\rho=0\) splits the space of quantum states into two, with the convex set of separable states fully contained in one half, typically chosen as \(\braket{\hat{W}}_\rho\geq0\).
Any state with \(\braket{\hat{W}}_\rho<0\) cannot be separable, and must hence be entangled.
In this way, one can use the measurement of \(\hat{W}\) as a tool to diagnose entangled states.
Though entanglement witnesses are by definition not able to detect all entangled states (those with \(\braket{\hat{W}}_\rho\geq0\)) and do not define an ordering relation between entangled states, they imply a significant resource economy:  entanglement witnessing entails the measurement of a (more or less complex) physical observable, in contrast to the knowledge of the full quantum state that is required for full-fledged entanglement measures such as von Neumann entropy or negativity \cite{guhne_toth_2009,friis_vitagliano_malik_huber_2018}.

Entanglement witnessing is well-established for systems with a tensor-product Hilbert space.
To develop an entanglement witness for LGTs, we need to identify observables that can faithfully divide out the separable states defined according to equation~\eqref{eq:gauge:separable}.
We consider witness operators of the form
\begin{equation}
    \hat{W}_\pm
    =\pm\left( \omega_\text{sep}^\pm\hat{\mathbb{1}}-\hat{C} \right)
    \,,
    \label{eq:witness:operator}
\end{equation}
with an observable \(\hat{C}=\sum_m \hat{O}^m_\lat{A} \hat{O}^m_\lat{B}\)  that is used to capture quantum correlations between subregions \(\lat{A}\) and \(\lat{B}\).

Most importantly, in the construction of \(\hat{C}\) we must employ only gauge invariant observables that preserve the local superselection rules, i.e., \(\hat{O}^m_\lat{A}\in\galg{A}\) and \(\hat{O}^m_\lat{B}\in\galg{B}\).
The constants \(\omega_\text{sep}^\pm\) are chosen to ensure a non--negative expectation value for all separable states.
By optimizing over all separable states, 
\begin{equation}
    \omega_\text{sep}^\pm
    = \underset{\rho_\text{sep}}{\max\!/\!\min} \braket{\hat{C}}_{\rho_\text{sep}}
    \,,
    \label{eq:witness:separable}
\end{equation}
the \(\braket{\hat{W}_\pm}_\rho=0\) hyperplane touches the boundary of the set of separable states, thus maximizing the entangled region detected by \(\hat{W}_\pm\) \cite{lewenstein_kraus_cirac_horodecki_2000}.
However, for the entanglement witness to be effective, at least some entangled state has to overcome the bounds from equation~\eqref{eq:witness:separable}, i.e., 
\begin{equation}
    \omega^- <\omega_\text{sep}^-\quad \mathrm{or}\quad \omega_\text{sep}^+ <\omega^+
    \label{eq:witness:condition}
\end{equation}
where \(\omega^\pm\) is the maximal/minimal eigenvalue of $\hat{C}$.

In practice, the direct sum in equation~\eqref{eq:gauge:separable} enables a substantial simplification of the optimization procedure in equation~\eqref{eq:witness:separable}.
Namely, we can solve the optimization procedure for product states within a fixed pair of compatible SSs \(\sss{A}\) and \(\sss{B}\),  
\begin{equation}
    \omega_\text{sep}^\pm(\sss{A},\sss{B})\!=\!
        \underset{\ket{\psi_\lat{A}},\ket{\psi_\lat{B}}}{\max\! /\! \min} 
    \sum_m 
        \bra{\psi_\lat{A}}\hat{O}^m_\lat{A}\ket{\psi_\lat{A}}\!
        \bra{\psi_\lat{B}}\hat{O}^m_\lat{B}\ket{\psi_\lat{B}}\,.
    \label{eq:witness:sector}
\end{equation}
Optimizing subsequently over all compatible pairs \(\sss{A}\) and \(\sss{B}\), one recovers \(\omega_\text{sep}^\pm\).
This procedure is significantly more efficient than performing a single optimization over the set of all separable states.
Moreover, in a scenario where there is additional information about the state being evaluated, a more sensitive witness may be obtained.
For instance, if the probability distribution \(p(\sss{A},\sss{B})\) is known, the constants
\begin{equation}
    \omega_\text{sep}^\pm (p)= \sum_{\left(\sss{A},\sss{B}\right)}p(\sss{A},\sss{B}) \omega_\text{sep}^{\pm}(\sss{A},\sss{B})
    \label{eq:witness:probability}
\end{equation}
yield tighter bounds that may detect entangled states that \(\omega_\text{sep}^\pm\) cannot.

In particular, if there is only a single fixed superselection pair for \(\lat{A}\) and \(\lat{B}\), the bounds from equation~\eqref{eq:witness:sector} can be used directly.

Notably, though we focus on the bipartite case, our separability criterion and the construction of the entanglement witness can be easily extended to the multipartite case.
Assume a scenario with subregions \(\mathcal{A}_1\dots\mathcal{A}_k\).
We can then define \(\hat{C}=\sum_m \hat{O}^m_{\lat{A}_1}\dots\hat{O}^m_{\lat{A}_k}\) (there are now more possibilities, since each summand can couple different subregions, i.e., some \(\hat{O}^m_{\lat{A}_q}\) can equal the identity in different combinations).
To calculate the analogues of equation~\eqref{eq:witness:sector}, one further needs to compute the compatible SS combinations \((\sssi{A}{1},\dots \sssi{A}{k})\), leading to bounds for multipartite separable states.

In the next sections, we provide concrete realizations of entanglement witnesses for \(\mathrm{U}(1)\) LGTs in 2+1D with explicit evaluations for equation~\eqref{eq:witness:separable}.
They provide additional details on the decomposition coming from equation~\eqref{eq:gauge:hilbert} and describe how the compatibility constraints between the SSs come about.

\section{Pure \texorpdfstring{\(\mathrm{U}(1)\)}{U(1)} gauge theory in 2+1D}
\label{sec:gauge}

To exemplify the above discussions, in this section we consider entanglement witnesses for a pure \(\mathrm{U}(1)\) LGT in 2+1D.
We will include dynamical matter in the next section.

The theory considered lives on a spatial lattice \(\lat{L}\), consisting of lattice sites \(j\), edges \(\ell\), and plaquettes \(p\).
We choose a two-dimensional local Hilbert space associated with each edge \(\ell\), spanned by \(\{\ket{0}_\ell,\ket{1}_\ell\}\) (see top left of figure~\ref{fg:main:lattice}) and whose local algebra is generated by the Pauli matrices \(\sigma_\ell\).
This choice allows us to find the following representations for the `electric field' operators \(\hat{E}_\ell\) and the \(\mathrm{U}(1)\) gauge generators \(g_j\):  
\begin{equation}
    \hat{E}_\ell= \frac{1}{2} \hat{\sigma}^z_\ell\,\text{ and }\,
    \hat{g}_j = \sum_{\ell\in\star j } \mathrm{sign}(\ell, j) \hat{E}_\ell \,,
    \label{eq:gauge:generator}
\end{equation}
where \(\star j\) denotes the set of edges whose boundary includes \(j\) and
\begin{equation}
    \mathrm{sign}(\ell,j) = 
    \begin{cases} 
    + 1 &\,\text{if}\quad j = \partial_0^+\ell\\
    -1  &\,\text{if}\quad j = \partial_0^-\ell
    \end{cases}
    \label{eq:gauge:orientation}
\end{equation}
accounts for the different relative orientations.
Under our conventions, for a horizontal (vertical) edge \(\ell\), \(\partial_0^+\ell\) is the left (bottom) site and \(\partial_0^-\ell\) is the right (upper) site.

Equation~\eqref{eq:gauge:generator} is a discretized version of the Gauss's law generator known from QED that is used, e.g., in the context of quantum link models \cite{chandrasekharan_wiese_1997,Wiese2013AnnPhys}.
Choosing a uniform and vanishing background of static charges, physical states \(\ket{\Psi}\) have to satisfy \(\hat{g}_j \ket{\Psi} = 0\) for every \(j \in \lat{L}\), see figure~\ref{fg:main:lattice} for an example.
Within the pictorial representation of figure~\ref{fg:main:lattice}, the physical states obey what is known in the literature as \textit{two-in-two-out rule}, i.e., at each vertex \(v\) two arrows are incoming and two are outgoing.

Within our scenario, the only physically meaningful operators \(\hat{O}\) are those that are gauge invariant, i.e., those with \([\hat{O},\hat{g}_j] = 0\) for every site \(j\).
Naturally, the set of gauge invariant operators includes the generators \(\hat{g}_j\) and electric fields \(\hat{E}_\ell\).
In addition to these, the present theory has gauge-invariant plaquette operators \(\hat{U}_p\) and \(\hat{U}_p^\dag\), 
\begin{equation}
    \hat{U}_p \, {=} \, \hat{\sigma}^-_{\ell_1}\otimes\hat{\sigma}^-_{\ell_2}\otimes
    \hat{\sigma}^+_{\ell_3}\otimes\hat{\sigma}^+_{\ell_4}\,\text{ and }\,
    \hat{U}_p^\dag \, {=} \, \hat{\sigma}^+_{\ell_1}\otimes\hat{\sigma}^+_{\ell_2}\otimes
    \hat{\sigma}^-_{\ell_3}\otimes\hat{\sigma}^-_{\ell_4} \,,
    \label{eq:gauge:plaquette}
\end{equation}
where the involved links are defined in figure~\ref{fg:main:lattice}.
Figure~\ref{fg:main:plaquette} depicts the action of the plaquette operators on gauge-invariant configurations.
It is possible to obtain further gauge invariant operators by linearly combining and/or considering tensor products among \(\hat{g}_j\), \(\hat{E}_\ell\), \(\hat{U}_p\), and \(\hat{U}_p^\dag\).
In particular, tensor products of plaquette operators on neighboring plaquettes yield gauge-invariant loop operators.

\subsection{Constructing a valid entanglement witness} 
\label{sec:gauge:bipartition}

We can define the set of boundary points \(\partial_0\lat{X}\) of a subregion \(\lat{X}\) as the points belonging to it that have at least one connected edge lying outside \(\lat{X}\).
For each site \(j \in \partial_0\lat{X}\), it is possible to associate an operator \(\hat{Z}_j\) that commutes with all locally gauge invariant operators of \(\lat{X}\),
\begin{equation}
    \hat{Z}_j
    = \sum_{\ell \in \lat{X}} \mathrm{sign}(\ell,j) \hat{E}_\ell
    \,.
    \label{eq:gauge:center}
\end{equation}
The superselection sector \(\hsss{X}\), with \(\sss{X} = \{Z_j\}_{j\in\partial_0\lat{X}}\), is the Hilbert space such that 
\begin{equation}
    \forall \ket{\Phi} \in \hsss{X}\,, \forall j \in \partial_0\lat{X}:\quad
    \hat{Z}_j\ket{\Phi} = Z_j\ket{\Phi}\,.
    \label{eq:gauge:sector}
\end{equation}
As mentioned before, an operator in \(\galg{X}\) cannot connect states belonging to different SSs, i.e., it does not couple \(\hsss[,1]{X}\) and \(\hsss[,2]{X}\) whenever \(\sss[,1]{X}\not= \sss[,2]{X}\). 
Given a concrete description of the generators of \(Z(\galg{X})\), there are some constraints that a label set \(Z\) has to fulfill in order to identify a physical superselection sector (see appendix~\ref{sec:appendix_A}).

With the above definitions, we are in the position to define a valid entanglement witness for the considered \(\mathrm{U}(1)\) gauge theory.  
Assume the two non-touching sub-regions \(\lat{A}\) and \(\lat{B}\) to contain \(N\) plaquettes each, as sketched in figure~\ref{fg:main}. 
We can then define a witness operator compatible with the prescriptions in section~\ref{sec:gauge:bipartition}  through 
\begin{equation}
    \hat{C} = \sum_{i=1}^{N}\plaqob{a_i} \otimes \plaqob{b_i}\,\text{ with }\,
    \plaqob{x_i} = \plaqop{x_i} e^{i\phi_{x_i}} + \plaqop{x_i}^\dag e^{-i\phi_{x_i}}
    \,.
    \label{eq:gauge:witness}
\end{equation}
Here, $a_i$ and $b_i$ label plaquettes in $\lat{A}$ and $\lat{B}$, respectively.  
The phases $\phi_{x_i}$ may be useful to optimize the witness depending on the physical situation, but, without restriction of generality, we choose $\phi_{x_i}=0$ in what follows.
Given that $\plaqop{a_i}$ and $\plaqop{b_i}$ are gauge invariant operators, it follows that $\hat{C}$ is a meaningful operator in the considered $\mathrm{U}(1)$ LGT.
In particular, $\hat{C}$ cannot change the border conditions of regions $\lat{A}$ and $\lat{B}$.
To evaluate $\braket{\hat{C}}$ and obtain the constants for equation~\eqref{eq:witness:condition}, we extract bounds for the expectation value of $\hat{C}$ both for generic states (not necessarily separable) and separable states.
Fixing a pair of SSs \(\hsss{A}\otimes\hsss{B}\), yields the bounds \(\omega_\mathrm{sep}^-(\sss{A},\sss{B})\) and \(\omega_\mathrm{sep}^+(\sss{A},\sss{B})\), from which we can define an entanglement witness via equation~\eqref{eq:witness:operator}.

\subsection{Concrete example}
\label{eq:ConcreteExampleU1}

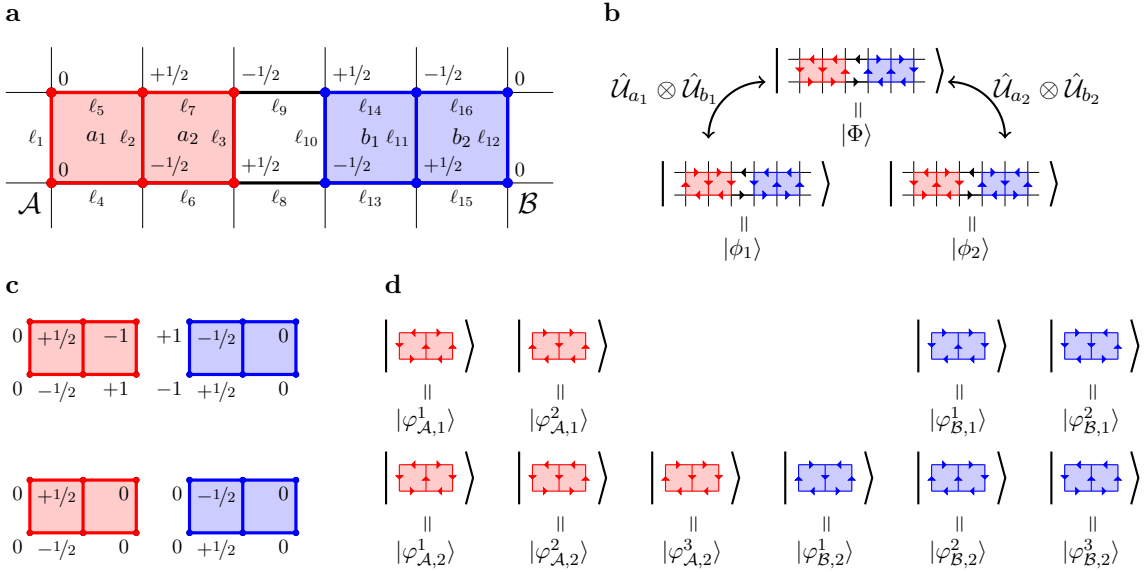
\begin{figure}[t]
    \centering
    \begin{subfigure}[t]{0.5\textwidth}
        \caption{}
        \begin{tikzpicture}[scale=1.2]
    \draw[fill, color= red!20] (0,0) rectangle (2,1);
    \draw[fill, color= blue!20] (3,0) rectangle (5,1);
    \draw[very thin] (-.5,-.5) grid (5.5,1.5);
    \draw[very thick] (0,0) grid (5,1);
    \draw[very thick,red] (0,0) grid (2,1);
    \draw[very thick, blue] (3,0) grid (5,1);
    \draw (0,0) node[anchor=north east]{$\mathcal{A}$};
    \draw (5,0) node[anchor=north west]{$\mathcal{B}$};
    \begin{scope}[every node/.style={scale=.65}]
        \draw (.5,.5) node[scale=1.2]{$a_1$};
        \draw (1.5,.5) node[scale=1.2]{$a_2$};
        \draw (3.5,.5) node[scale=1.2]{$b_1$};
        \draw (4.5,.5) node[scale=1.2]{$b_2$};
    
        \draw (0,.5) node[anchor=east]{$\ell_1$};
        \draw (1,.5) node[anchor=east]{$\ell_2$};
        \draw (2,.5) node[anchor=east]{$\ell_3$};
        \draw (.5,0) node[anchor=north]{$\ell_4$};
        \draw (.5,1) node[anchor=north]{$\ell_5$};
        \draw (1.5,0) node[anchor=north]{$\ell_6$};
        \draw (1.5,1) node[anchor=north]{$\ell_7$};
        \draw (2.5,0) node[anchor=north]{$\ell_8$};
        \draw (2.5,1) node[anchor=north]{$\ell_9$};
        \draw (3,.5) node[anchor = east]{$\ell_{10}$};
        \draw (4,.5) node[anchor = east]{$\ell_{11}$};
        \draw (5,.5) node[anchor = east]{$\ell_{12}$};
        \draw (3.5,0) node[anchor = north]{$\ell_{13}$};
        \draw (3.5,1) node[anchor = north]{$\ell_{14}$};
        \draw (4.5,0) node[anchor = north]{$\ell_{15}$};
        \draw (4.5,1) node[anchor = north]{$\ell_{16}$};
    \end{scope}
    \begin{scope}[every node/.style={scale=.7}]
        \draw[fill] (0,0) circle (.05) node[anchor=south west]{$0$};
        \draw[fill] (1,0) circle (.05) node[anchor=south west]{$-\nicefrac{1}{2}$};
        \draw[fill] (2,0) circle (.05) node[anchor=south west]{$+\nicefrac{1}{2}$};
        \draw[fill] (3,0) circle (.05) node[anchor=south west]{$-\nicefrac{1}{2}$};
        \draw[fill] (4,0) circle (.05) node[anchor=south west]{$+\nicefrac{1}{2}$};
        \draw[fill] (5,0) circle (.05) node[anchor=south west]{$0$};
        \draw[fill] (5,1) circle (.05) node[anchor=south west]{$0$};
        \draw[fill] (4,1) circle (.05) node[anchor=south west]{$-\nicefrac{1}{2}$};
        \draw[fill] (3,1) circle (.05) node[anchor=south west]{$+\nicefrac{1}{2}$};
        \draw[fill] (2,1) circle (.05) node[anchor=south west]{$-\nicefrac{1}{2}$};
        \draw[fill] (1,1) circle (.05) node[anchor=south west]{$+\nicefrac{1}{2}$};
        \draw[fill] (0,1) circle (.05) node[anchor=south west]{$0$};
        \foreach \i in {0,...,2} \foreach \j in {0,1}
            \draw[fill, red] (\i,\j) circle (.05);
        \foreach \i in {3,...,5} \foreach \j in {0,1}
            \draw[fill, blue] (\i,\j) circle (.05);

    \end{scope}
\end{tikzpicture}
        \label{fg:gauge:example}
    \end{subfigure}
    \hfill
    \begin{subfigure}[t]{0.48\textwidth}
        \caption{}
        \begin{tikzpicture}[scale=.3]
    \draw[<->, thick] (1.5,2.5) arc (180:90:2.5) node[anchor=east,pos=0.45,scale=.9,xshift=+1mm,yshift=+2mm]
        {$\hat{\mathcal{U}}_{a_1}\otimes\hat{\mathcal{U}}_{b_1}$};
    \draw[<->, thick] (14.5,2.5) arc (0:90:2.5) node[anchor=west,pos=.45,scale=.9,,xshift=-1mm,yshift=+2mm]
        {$\hat{\mathcal{U}}_{a_2}\otimes\hat{\mathcal{U}}_{b_2}$};
    \begin{tzket}[shift={(0,0)}]
        \begin{scope}[shift={(-.5,0)}]
            \draw[very thin] (0.5,-.5) grid (6.5,1.5);
        \end{scope}            
        \draw[shift={(0.5,0)},fill,color=red!20] (0,0) rectangle (2,1);
        \begin{scope}[shift={(0.5,0)},red,every node/.style={sloped,allow upside down,scale=.4}]
            \draw (0,0) -- node{\midarrow} (0,1);
            \draw (1,1) -- node{\midarrow} (1,0);
            \draw (2,1) -- node{\midarrow} (2,0);
            \draw (1,0) -- node{\midarrow} (0,0);
            \draw (2,0) -- node{\midarrow} (1,0);
            \draw (0,1) -- node{\midarrow} (1,1);
            \draw (1,1) -- node{\midarrow} (2,1);
        \end{scope}
        \begin{scope}[shift={(2.5,0)},every node/.style={sloped,allow upside down,scale=.4}]
            \draw (0,0) -- node{\midarrow} (1,0);
            \draw (1,1) -- node{\midarrow} (0,1);
        \end{scope}
        \draw[shift={(3.5,0)},fill,color=blue!20] (0,0) rectangle (2,1);
        \begin{scope}[shift={(3.5,0)},blue,every node/.style={sloped,allow upside down,scale=.4}]
            \draw (0,1) -- node{\midarrow} (0,0);
            \draw (1,0) -- node{\midarrow} (1,1);
            \draw (2,0) -- node{\midarrow} (2,1);
            \draw (0,0) -- node{\midarrow} (1,0);
            \draw (1,1) -- node{\midarrow} (0,1);
            \draw (1,0) -- node{\midarrow} (2,0);
            \draw (2,1) -- node{\midarrow} (1,1);
        \end{scope}
        \tzwidth{6}
        \draw (3,-.5) node[anchor=north,scale=.8]{\rotatebox{270}{$=$}};
        \draw (3,-1.5) node[anchor=north,scale=.8]{$ \ket{\phi_1}$};
    \end{tzket}
    \begin{tzket}[shift={(5,5)}]
        \begin{scope}[shift={(-.5,0)}]
            \draw[very thin] (0.5,-.5) grid (6.5,1.5);
        \end{scope}            
        \draw[shift={(0.5,0)},fill,color=red!20] (0,0) rectangle (2,1);
        \begin{scope}[shift={(0.5,0)},red,every node/.style={sloped,allow upside down,scale=.4}]
            \draw (0,1) -- node{\midarrow} (0,0);
            \draw (1,1) -- node{\midarrow} (1,0);
            \draw (2,0) -- node{\midarrow} (2,1);
            \draw (0,0) -- node{\midarrow} (1,0);
            \draw (1,0) -- node{\midarrow} (2,0);
            \draw (1,1) -- node{\midarrow} (0,1);
            \draw (2,1) -- node{\midarrow} (1,1);
        \end{scope}
        \begin{scope}[shift={(2.5,0)},every node/.style={sloped,allow upside down,scale=.4}]
            \draw (0,0) -- node{\midarrow} (1,0);
            \draw (1,1) -- node{\midarrow} (0,1);
        \end{scope}
        \draw[shift={(3.5,0)},fill,color=blue!20] (0,0) rectangle (2,1);
        \begin{scope}[shift={(3.5,0)},blue,every node/.style={sloped,allow upside down,scale=.4}]
            \draw (0,0) -- node{\midarrow} (0,1);
            \draw (1,0) -- node{\midarrow} (1,1);
            \draw (2,1) -- node{\midarrow} (2,0);
            \draw (1,0) -- node{\midarrow} (0,0);
            \draw (0,1) -- node{\midarrow} (1,1);
            \draw (2,0) -- node{\midarrow} (1,0);
            \draw (1,1) -- node{\midarrow} (2,1);
        \end{scope}
        \tzwidth{6}
        \draw (3,-.5) node[anchor=north,scale=.8]{\rotatebox{270}{$=$}};
        \draw (3,-1.5) node[anchor=north,scale=.8]{$ \ket{\Phi}$};
    \end{tzket}
    \begin{tzket}[shift={(10,0)}]
        \begin{scope}[shift={(-.5,0)}]
            \draw[very thin] (0.5,-.5) grid (6.5,1.5);
        \end{scope}            
        \draw[shift={(0.5,0)},fill,color=red!20] (0,0) rectangle (2,1);
        \begin{scope}[shift={(0.5,0)},red,every node/.style={sloped,allow upside down,scale=.4}]
            \draw (0,1) -- node{\midarrow} (0,0);
            \draw (1,0) -- node{\midarrow} (1,1);
            \draw (2,1) -- node{\midarrow} (2,0);
            \draw (0,0) -- node{\midarrow} (1,0);
            \draw (2,0) -- node{\midarrow} (1,0);
            \draw (1,1) -- node{\midarrow} (0,1);
            \draw (1,1) -- node{\midarrow} (2,1);
        \end{scope}
        \begin{scope}[shift={(2.5,0)},every node/.style={sloped,allow upside down,scale=.4}]
            \draw (0,0) -- node{\midarrow} (1,0);
            \draw (1,1) -- node{\midarrow} (0,1);
        \end{scope}
        \draw[shift={(3.5,0)},fill,color=blue!20] (0,0) rectangle (2,1);
        \begin{scope}[shift={(3.5,0)},blue,every node/.style={sloped,allow upside down,scale=.4}]
            \draw (0,0) -- node{\midarrow} (0,1);
            \draw (1,1) -- node{\midarrow} (1,0);
            \draw (2,0) -- node{\midarrow} (2,1);
            \draw (1,0) -- node{\midarrow} (0,0);
            \draw (0,1) -- node{\midarrow} (1,1);
            \draw (1,0) -- node{\midarrow} (2,0);
            \draw (2,1) -- node{\midarrow} (1,1);
        \end{scope}
        \tzwidth{6}
        \draw (3,-.5) node[anchor=north,scale=.8]{\rotatebox{270}{$=$}};
        \draw (3,-1.5) node[anchor=north,scale=.8]{$ \ket{\phi_2}$};
    \end{tzket}
\end{tikzpicture}
        \label{fg:gauge:entangled}
    \end{subfigure}
    \vfill
    \begin{subfigure}[t]{0.3\textwidth}
        \caption{}
        \begin{tikzpicture}[scale=0.7]
    \begin{scope}[every node/.style={scale=.7}]
        \draw[fill, color = red!20] (0,0) rectangle (2,1);
        \draw[fill, color = blue!20] (3,0) rectangle (5,1);
        \draw[very thick,red] (0,0) grid (2,1);
        \draw[very thick,blue] (3,0) grid (5,1);

        \draw[fill] (0,0) circle (.05) node[anchor=north east]{$0$};
        \draw[fill] (1,0) circle (.05) node[anchor=north east]{$-\nicefrac{1}{2}$};
        \draw[fill] (2,0) circle (.05) node[anchor=north east]{$+1$};
        \draw[fill] (2,1) circle (.05) node[anchor=north east]{$-1$};
        \draw[fill] (1,1) circle (.05) node[anchor=north east]{$+\nicefrac{1}{2}$};
        \draw[fill] (0,1) circle (.05) node[anchor=north east]{$0$};
        
        \draw[fill] (3,0) circle (.05) node[anchor = north east]{$-1$};
        \draw[fill] (4,0) circle (.05) node[anchor = north east]{$+\nicefrac{1}{2}$};
        \draw[fill] (5,0) circle (.05) node[anchor = north east]{$0$};
        \draw[fill] (5,1) circle (.05) node[anchor = north east]{$0$};
        \draw[fill] (4,1) circle (.05) node[anchor = north east]{$-\nicefrac{1}{2}$};
        \draw[fill] (3,1) circle (.05) node[anchor = north east]{$+1$};
        \foreach \i in {0,...,2} \foreach \j in {0,1}
            \draw[fill, red] (\i,\j) circle (.05);
        \foreach \i in {3,...,5} \foreach \j in {0,1}
            \draw[fill, blue] (\i,\j) circle (.05);
    \end{scope}
    \begin{scope}[shift={(0,-3)},every node/.style={scale=.7}]
        \draw[fill, color = red!20] (0,0) rectangle (2,1);
        \draw[fill, color = blue!20] (3,0) rectangle (5,1);
        \draw[very thick,red] (0,0) grid (2,1);
        \draw[very thick,blue] (3,0) grid (5,1);

        \draw[fill] (0,0) circle (.05) node[anchor=north east]{$0$};
        \draw[fill] (1,0) circle (.05) node[anchor=north east]{$-\nicefrac{1}{2}$};
        \draw[fill] (2,0) circle (.05) node[anchor=north east]{$0$};
        \draw[fill] (2,1) circle (.05) node[anchor=north east]{$0$};
        \draw[fill] (1,1) circle (.05) node[anchor=north east]{$+\nicefrac{1}{2}$};
        \draw[fill] (0,1) circle (.05) node[anchor=north east]{$0$};
        
        \draw[fill] (3,0) circle (.05) node[anchor = north east]{$0$};
        \draw[fill] (4,0) circle (.05) node[anchor = north east]{$+\nicefrac{1}{2}$};
        \draw[fill] (5,0) circle (.05) node[anchor = north east]{$0$};
        \draw[fill] (5,1) circle (.05) node[anchor = north east]{$0$};
        \draw[fill] (4,1) circle (.05) node[anchor = north east]{$-\nicefrac{1}{2}$};
        \draw[fill] (3,1) circle (.05) node[anchor = north east]{$0$};
        \foreach \i in {0,...,2} \foreach \j in {0,1}
            \draw[fill, red] (\i,\j) circle (.05);
        \foreach \i in {3,...,5} \foreach \j in {0,1}
            \draw[fill, blue] (\i,\j) circle (.05);
    \end{scope}
\end{tikzpicture}
        \label{fg:gauge:sectors}
    \end{subfigure}
    \hfill
    \begin{subfigure}[t]{0.67\textwidth}
        \caption{}
        \begin{tikzpicture}[scale=.35]
    \def\xshift{5}
    \def\yshift{-5}
    \begin{tzket}
        \draw[fill,color=red!20] (0,0) rectangle (2,1);
        \begin{scope}[red, every node/.style={sloped,allow upside down,scale=0.35}]
            \draw (0,0) -- node{\midarrow} (1,0);
            \draw (1,0) -- node{\midarrow} (1,1);
            \draw (1,1) -- node{\midarrow} (0,1);
            \draw (0,1) -- node{\midarrow} (0,0);
            \draw (1,1) -- node{\midarrow} (2,1);
            \draw (2,0) -- node{\midarrow} (2,1);
            \draw (2,0) -- node{\midarrow} (1,0);
        \end{scope}
        \tzwidth{2}
        \draw (1,-.6) node[anchor=north,scale=.8]{\rotatebox{-90}{$=$}};
        \draw (1,-1.5) node[anchor=north,scale=.8]{$\ket{\varphi_{\lat{A},1}^1}$};
    \end{tzket}
    \begin{tzket}[shift={(\xshift,0)}]
        \draw[fill,color=red!20] (0,0) rectangle (2,1);
        \begin{scope}[red, every node/.style={sloped,allow upside down,scale=0.35}]
            \draw (1,0) -- node{\midarrow} (0,0);
            \draw (1,1) -- node{\midarrow} (1,0);
            \draw (0,1) -- node{\midarrow} (1,1);
            \draw (0,0) -- node{\midarrow} (0,1);
            \draw (1,1) -- node{\midarrow} (2,1);
            \draw (2,0) -- node{\midarrow} (2,1);
            \draw (2,0) -- node{\midarrow} (1,0);
        \end{scope}
        \tzwidth{2}
        \draw (1,-.6) node[anchor=north,scale=.8]{\rotatebox{-90}{$=$}};
        \draw (1,-1.5) node[anchor=north,scale=.8]{$\ket{\varphi_{\lat{A},1}^2}$};
    \end{tzket}
    \begin{tzket}[shift={(0,\yshift)}]
        \draw[fill,color=red!20] (0,0) rectangle (2,1);
        \begin{scope}[red, every node/.style={sloped,allow upside down,scale=0.35}]
            \draw (0,0) -- node{\midarrow} (1,0);
            \draw (1,0) -- node{\midarrow} (1,1);
            \draw (1,1) -- node{\midarrow} (0,1);
            \draw (0,1) -- node{\midarrow} (0,0);
            \draw (1,1) -- node{\midarrow} (2,1);
            \draw (2,1) -- node{\midarrow} (2,0);
            \draw (2,0) -- node{\midarrow} (1,0);
        \end{scope}
        \tzwidth{2}
        \draw (1,-.6) node[anchor=north,scale=.8]{\rotatebox{-90}{$=$}};
        \draw (1,-1.5) node[anchor=north,scale=.8]{$\ket{\varphi_{\lat{A},2}^1}$};
    \end{tzket}
    \begin{tzket}[shift={(\xshift,\yshift)}]
        \draw[fill,color=red!20] (0,0) rectangle (2,1);
        \begin{scope}[red, every node/.style={sloped,allow upside down,scale=0.35}]
            \draw (0,0) -- node{\midarrow} (1,0);
            \draw (1,1) -- node{\midarrow} (1,0);
            \draw (1,1) -- node{\midarrow} (0,1);
            \draw (0,1) -- node{\midarrow} (0,0);
            \draw (2,1) -- node{\midarrow} (1,1);
            \draw (2,0) -- node{\midarrow} (2,1);
            \draw (1,0) -- node{\midarrow} (2,0);
        \end{scope}
        \tzwidth{2}
        \draw (1,-.6) node[anchor=north,scale=.8]{\rotatebox{-90}{$=$}};
        \draw (1,-1.5) node[anchor=north,scale=.8]{$\ket{\varphi_{\lat{A},2}^2}$};
    \end{tzket}
    \begin{tzket}[shift={(2*\xshift,\yshift)}]
        \draw[fill,color=red!20] (0,0) rectangle (2,1);
        \begin{scope}[red, every node/.style={sloped,allow upside down,scale=0.35}]
            \draw (1,0) -- node{\midarrow} (0,0);
            \draw (1,1) -- node{\midarrow} (1,0);
            \draw (0,1) -- node{\midarrow} (1,1);
            \draw (0,0) -- node{\midarrow} (0,1);
            \draw (1,1) -- node{\midarrow} (2,1);
            \draw (2,1) -- node{\midarrow} (2,0);
            \draw (2,0) -- node{\midarrow} (1,0);
        \end{scope}
        \tzwidth{2}
        \draw (1,-.6) node[anchor=north,scale=.8]{\rotatebox{-90}{$=$}};
        \draw (1,-1.5) node[anchor=north,scale=.8]{$\ket{\varphi_{\lat{A},2}^3}$};
    \end{tzket}
    \begin{tzket}[shift={(4*\xshift,0)}]
        \draw[fill,color=blue!20] (0,0) rectangle (2,1);
        \begin{scope}[blue, every node/.style={sloped,allow upside down,scale=0.35}]
                \draw (0,1) -- node{\midarrow} (0,0);
                \draw (1,0) -- node{\midarrow} (0,0);
                \draw (0,1) -- node{\midarrow} (1,1);
                \draw (1,0) -- node{\midarrow} (1,1);
                \draw (1,1) -- node{\midarrow} (2,1);
                \draw (2,1) -- node{\midarrow} (2,0);
                \draw (2,0) -- node{\midarrow} (1,0);
        \end{scope}
        \tzwidth{2}
        \draw (1,-.6) node[anchor=north,scale=.8]{\rotatebox{-90}{$=$}};
        \draw (1,-1.5) node[anchor=north,scale=.8]{$\ket{\varphi_{\lat{B},1}^1}$};
    \end{tzket}
    \begin{tzket}[shift={(5*\xshift,0)}]
        \draw[fill,color=blue!20] (0,0) rectangle (2,1);
        \begin{scope}[blue, every node/.style={sloped,allow upside down,scale=0.35}]
                \draw (0,1) -- node{\midarrow} (0,0);
                \draw (1,0) -- node{\midarrow} (0,0);
                \draw (0,1) -- node{\midarrow} (1,1);
                \draw (1,1) -- node{\midarrow} (1,0);
                \draw (2,1) -- node{\midarrow} (1,1);
                \draw (2,0) -- node{\midarrow} (2,1);
                \draw (1,0) -- node{\midarrow} (2,0);
        \end{scope}
        \tzwidth{2}
        \draw (1,-.6) node[anchor=north,scale=.8]{\rotatebox{-90}{$=$}};
        \draw (1,-1.5) node[anchor=north,scale=.8]{$\ket{\varphi_{\lat{B},1}^2}$};
    \end{tzket}
    \begin{tzket}[shift={(3*\xshift,\yshift)}]
        \draw[fill,color=blue!20] (0,0) rectangle (2,1);
        \begin{scope}[blue, every node/.style={sloped,allow upside down,scale=0.35}]
            \draw (1,0) -- node{\midarrow} (0,0);
            \draw (1,1) -- node{\midarrow} (1,0);
            \draw (0,1) -- node{\midarrow} (1,1);
            \draw (0,0) -- node{\midarrow} (0,1);
            \draw (2,1) -- node{\midarrow} (1,1);
            \draw (2,0) -- node{\midarrow} (2,1);
            \draw (1,0) -- node{\midarrow} (2,0);
        \end{scope}
        \tzwidth{2}
        \draw (1,-.6) node[anchor=north,scale=.8]{\rotatebox{-90}{$=$}};
        \draw (1,-1.5) node[anchor=north,scale=.8]{$\ket{\varphi_{\lat{B},2}^1}$};
    \end{tzket}
    \begin{tzket}[shift={(4*\xshift,\yshift)}]
        \draw[fill,color=blue!20] (0,0) rectangle (2,1);
        \begin{scope}[blue, every node/.style={sloped,allow upside down,scale=0.35}]
            \draw (1,0) -- node{\midarrow} (0,0);
            \draw (1,0) -- node{\midarrow} (1,1);
            \draw (0,1) -- node{\midarrow} (1,1);
            \draw (0,0) -- node{\midarrow} (0,1);
            \draw (1,1) -- node{\midarrow} (2,1);
            \draw (2,1) -- node{\midarrow} (2,0);
            \draw (2,0) -- node{\midarrow} (1,0);
        \end{scope}
        \tzwidth{2}
        \draw (1,-.6) node[anchor=north,scale=.8]{\rotatebox{-90}{$=$}};
        \draw (1,-1.5) node[anchor=north,scale=.8]{$\ket{\varphi_{\lat{B},2}^2}$};
    \end{tzket}
    \begin{tzket}[shift={(5*\xshift,\yshift)}]
        \draw[fill,color=blue!20] (0,0) rectangle (2,1);
        \begin{scope}[blue, every node/.style={sloped,allow upside down,scale=0.35}]
            \draw (0,0) -- node{\midarrow} (1,0);
            \draw (1,0) -- node{\midarrow} (1,1);
            \draw (1,1) -- node{\midarrow} (0,1);
            \draw (0,1) -- node{\midarrow} (0,0);
            \draw (2,1) -- node{\midarrow} (1,1);
            \draw (2,0) -- node{\midarrow} (2,1);
            \draw (1,0) -- node{\midarrow} (2,0);
        \end{scope}
        \tzwidth{2}
        \draw (1,-.6) node[anchor=north,scale=.8]{\rotatebox{-90}{$=$}};
        \draw (1,-1.5) node[anchor=north,scale=.8]{$\ket{\varphi_{\lat{B},2}^3}$};
    \end{tzket}
\end{tikzpicture}
        \label{fg:gauge:bases}
    \end{subfigure}
    \caption{\textbf{(a)} Example geometry and SS considered in section~\ref{sec:gauge}, with $N=2$ plaquettes in both subregions \(\lat{A}\) and  \(\lat{B}\).
        \textbf{(b)} Constituents from which the entangled state $\ket{\xi}$ of the pure-gauge $\mathrm{U}(1)$ theory is constructed and action of the plaquette-operator components of the operator $\hat{C}$ defined in Eq.~\eqref{eq:gauge:2pwitness}.
        \textbf{(c)} After tracing out the region neither in $\lat{A}$ nor $\lat{B}$ (i.e., the two central bonds in panel a), 
one obtains two pairs of allowed SSs.
    Top: \(\hsss[,1]{A} \otimes \hsss[,1]{B}\).
    Bottom: \(\hsss[,2]{A} \otimes \hsss[,2]{B}\).
    \textbf{(d)} Basis states associated with $\hsss[,1]{A}$ (top left, red), $\hsss[,1]{B}$ (top right, blue), $\hsss[,2]{A}$ (bottom left, red), and  $\hsss[,2]{B}$ (bottom right, blue). 
} 
    \label{fg:gauge}
\end{figure}

To give an example how the witness defined through equation~\eqref{eq:gauge:witness} works, we consider the simple geometry and associated SS depicted in figure~\ref{fg:gauge:example} (in appendix~\ref{sec:appendix_B}, we illustrate a \textit{reduction scheme} that permits to increase the efficiency in dealing with larger regions).
In this scenario, equation~\eqref{eq:gauge:witness} becomes
\begin{equation}
    \hat{C} = 
        \plaqob{a_1} \otimes\plaqob{b_1} +
        \plaqob{a_2} \otimes\plaqob{b_2}\,,
    \label{eq:gauge:2pwitness}
\end{equation}
with $\omega^- = -\sqrt{2}$ and $\omega^+ = +\sqrt{2}$. 

There are two pairs of SSs associated with regions \(\lat{A}\) and \(\lat{B}\) compatible with the SS of the larger region \(\lat{L}\), sketched in figure~\ref{fg:gauge:sectors}.
We label them \(\left(\sss[,1]{A},\sss[,1]{B}\right)\), top of figure~\ref{fg:gauge:sectors}, and \(\left(\sss[,2]{A},\sss[,2]{B}\right)\), bottom of figure~\ref{fg:gauge:sectors}. 
Using fixed basis states for \(\hsss[,1]{A}\otimes\hsss[,1]{B}\) and \(\hsss[,2]{A}\otimes\hsss[,2]{B}\), depicted in figure~\ref{fg:gauge:bases}, we explicitly evaluate the action of \(\hat{C}\) on separable states between regions \(\lat{A}\) and \(\lat{B}\).
The first case is relatively simple because \(\hat{C}\ket{\varphi^i_{\lat{A},1}}\otimes \ket{\varphi^j_{\lat{B},1}}=0\) for all \(i,j=1,2\).
This is the case since \(\hat{C}\), as defined in equation~\eqref{eq:gauge:witness}, can only combine a flippable plaquette in \(\lat{A}\) with an unflippable one in \(\lat{B}\) or vice versa. 
Thus, \(\hat{C}|_{\hsss[,1]{A}\otimes\hsss[,1]{B}}= 0\) and this operator cannot witness entanglement in this SS.

The second case is less immediate, and requires us to explicitly evaluate \(\braket{\hat{C}}\) for an arbitrary product state in \(\hsss[,2]{A}\otimes\hsss[,2]{B}\) and then optimize it to extract \(\omega_\mathrm{sep}^-(\sss[,2]{A},\sss[,2]{B})\) and \(\omega_\mathrm{sep}^+(\sss[,2]{A},\sss[,2]{B})\).
Consider a product state \(\ket{\Psi_\mathrm{prod}} = \ket{\Psi_{\lat{A}}}\otimes \ket{\Psi_{\lat{B}}}\in\hsss[,1]{A}\otimes\hsss[,1]{B}\) and an expansion in basis states as
\begin{align}
    \ket{\Psi_{\lat{A}}} = \alpha_1 \ket{\varphi_{\lat{A},2}^1} + \alpha_2 \ket{\varphi_{\lat{A},2}^2} + \alpha_3 \ket{\varphi_{\lat{A},2}^3}\,\text{ and }\,
    \label{eq:gauge:decomposition}
    \ket{\Psi_{\lat{B}}} = \beta_1 \ket{\varphi_{\lat{B},2}^1} + \beta_2 \ket{\varphi_{\lat{B},2}^2} + \beta_3 \ket{\varphi_{\lat{B},2}^3} \,.
\end{align}  
Then we have (see figure~\ref{fg:main:plaquette})
\begin{align}
    \langle\hat{C}\rangle_{\mathrm{sep}}&= (\alpha_1^*\alpha_3 + \mbox{ c.c.})(\beta_1^*\beta_3 + \mbox{ c.c.}) + (\alpha_1^*\alpha_2 + \mbox{ c.c.}) (\beta_1^*\beta_2 + \mbox{ c.c.})\,.
    \label{eq:gauge:expvalue}
\end{align}
Setting \(\alpha_i = a_i e^{i\phi_i}\) and \(\beta_i = b_i e^{i\eta_i}\), and choosing without restriction of generality \(\phi_1=\eta_1=0\), we can rewrite equation~\eqref{eq:gauge:expvalue} as
\begin{align}
    \langle\hat{C}\rangle_{\mathrm{sep}} =\, & 
    4 a_1 b_1 \sum_{i=2}^3a_ib_i\cos\phi_i \cos{\eta_i}\,.
    \label{eq:gauge:expvalue2}
\end{align}
Using that \(\cos\phi_i \cos{\eta_i}\) takes values in \([-1,+1]\), we can optimize \(\pm 4 a_1b_1(a_2b_2 + a_3 b_3)\) under the constraints given by the normalization of \(\ket{\Psi_{\lat{A}}}\) and \(\ket{\Psi_{\lat{B}}}\) (i.e., \(\sum_i a_i^2 = 1\) and \(\sum_i b_i^2 = 1\)).
As a consequence, we find 
\begin{equation}
    -1\leq \bra{\Psi_\mathrm{prod}}\hat{{C}}\ket{\Psi_\mathrm{prod}} \leq 1\,.
    \label{eq:gauge:bounds}
\end{equation}
The minimum is achieved, e.g., for \(\ket{\Psi_\mathrm{prod}} = \frac{1}{2}(\ket{\varphi^1_{\lat{A},2}} + \ket{\varphi^2_{\lat{A},2}})\otimes (\ket{\varphi^1_{\lat{B},2}}-\ket{\varphi^2_{\lat{B},2}})\), while the maximum is reached, e.g., with \(\ket{\Psi_\mathrm{prod}} =\frac{1}{2}(\ket{\varphi^1_{\lat{A},2} + \ket{\varphi^2_{\lat{A},2}}})\otimes (\ket{\varphi^1_{\lat{B},2}}+\ket{\varphi^2_{\lat{B},2}})\). 
Since equation~\eqref{eq:gauge:expvalue} is invariant under the exchange of \(\alpha_3\) with \(\alpha_2\) and of \(\beta_2\) with \(\beta_3\), we can recover an entire class of separable states that give extremal expectation values of \(\hat{C}\). 

We can now combine the above bounds to define entanglement witnesses for a generic \(\ket{\Psi_\lat{L}} \in \hsss{L}\). 
Assuming there is no information about the reduction to the subregion \(\lat{A}\cup\lat{B}\), the reduced state might be a mixture of states of \(\hsss[,1]{A}\otimes\hsss[,1]{B}\) and \(\hsss[,2]{A}\otimes\hsss[,2]{B}\), so we employ equation~\eqref{eq:witness:sector} and optimize over both pairs of SSs. 
Hence, it follows that the optimal bounds are given by \(\omega_\mathrm{sep}^-=\min_{j=1,2}\omega_\mathrm{sep}^-(\sss[,j]{A},\sss[,j]{B})=-1\) and  \(\omega_\mathrm{sep}^+=\max_{j=1,2}\omega_\mathrm{sep}^+(\sss[,j]{A},\sss[,j]{B})=1\) and the optimal entanglement witnesses are 
\begin{equation}
    \hat{W}_+ =\hat{\mathbb{1}} - \hat{C} \text{ and } \hat{W}_- = \hat{\mathbb{1}} + \hat{C} \,.
    \label{eq:gauge:optwitnesses}
\end{equation}
Given that the general bounds are \(\omega^- = -\sqrt{2}\) and \(\omega^+ = +\sqrt{2}\) reported below Eq.~\eqref{eq:gauge:2pwitness}, we conclude that any state with expectation value 
\(1-\sqrt{2}\leq \braket{\hat{{W}}_\pm}<0\) is detected as entangled. 

An example of an entangled state between regions \(\lat{A}\) and \(\lat{B}\) that saturates these bounds is \(\ket{\xi} = \frac 12 \ket{\phi_1} +  \frac{1}{\sqrt{2}}\ket{\Phi}  + \frac 12 \ket{\phi_2}\), with \(\ket{\phi_i}\) and \(\ket{\Phi}\) defined in Fig.~\ref{fg:gauge:entangled}.
Using that \(\hat{C}\ket{\Phi} = \ket{\phi_1} + \ket{\phi_2}\) and \(\hat{C}\ket{\phi_1} = \hat{C}\ket{\phi_2}  = \ket{\Phi}\), we obtain \(\hat{C}\ket{\xi}= \sqrt{2}\ket{\xi}\).
Thus, \( \braket{\hat{W}_+}_{\xi}=1 - \sqrt{2}\), correctly revealing that \(\ket{\xi}\) is entangled.

\section{Fermionic matter coupled to \texorpdfstring{\(\mathrm{U}(1)\)}{U(1)} gauge field}
\label{sec:matter}

Until this point, we have considered a pure gauge theory.
In this section, we generalize the discussion to include dynamical matter.

\subsection{Notations}

As an illustrative example, we consider---in addition to gauge fields living on the links the lattice \(\lat{L}\)---single-species fermions living on the lattice sites.
Concretely, we consider staggered fermions~\cite{kogut_1979} (see, e.g., references~\cite{Wiese2013AnnPhys,Hashizume2021ArXiv} for applications to quantum link models), 
where negatively (positively) charged (anti-)particles are associated in an alternating fashion to the sites of the lattice.
The gauge generators of equation~\eqref{eq:gauge:generator} are then modified to 
\begin{equation}
    \hat{g}_j = \hat{\psi}^\dag_j \hat{\psi}_j - \frac{1-(-1)^{(x_j+y_j)}}{2} - \sum_{\ell} \mathrm{sign}(\ell,j)\hat{E}_\ell\,,
    \label{eq:matter:generator}
\end{equation}
where \(\hat{\psi}_j\) (\(\hat{\psi}^\dag_j\)) are fermionic annihilation (creation) operators associated with lattice site \(j\).

\begin{figure}[t]
    \centering
    \begin{subfigure}[t]{0.34\textwidth}
        \caption{)}
        \begin{tikzpicture}[scale=1.2]  
        \useasboundingbox (-.5,-1.5) rectangle +(9,3);
        \draw[fill,color=red!20] (0,0) rectangle (1,1);
        \draw[fill,color=blue!20] (2,0) rectangle (3,1);
        \draw[very thin] (-.5,-.5) grid (3.5,1.5);
        \draw[thick] (0,0) grid (3,1);
        \draw (.5,.5) node[scale=.9]{${a_1}$};
        \draw (2.5,.5) node[scale=.9]{${b_1}$};
        \begin{scope}[every  node/.style={anchor=north east,scale=.6}]
            \draw[fill] (0,0) circle (.05) node{$-1$};
            \draw[fill] (1,0) circle (.05) node{$+\nicefrac{1}{2}$};
            \draw[fill] (2,0) circle (.05) node{$-\nicefrac{1}{2}$};
            \draw[fill] (3,0) circle (.05) node{$0$};
        \end{scope}   
        \begin{scope}[every  node/.style={anchor=south west,scale=.6}]
            \draw[fill] (3,1) circle (.05) node{$0$};
            \draw[fill] (2,1) circle (.05) node{$+\nicefrac{1}{2}$};
            \draw[fill] (1,1) circle (.05) node{$-\nicefrac{1}{2}$};
            \draw[fill] (0,1) circle (.05) node{$0$};
        \end{scope} 
        \begin{scope}[every node/.style ={scale=.8,pos=.5}]
            \draw (0,0) -- (0,1) node[anchor=east]{$\ell_1$};
            \draw (1,0) -- (1,1) node[anchor=west]{$\ell_2$};
            \draw (0,0) -- (1,0) node[anchor=north]{$\ell_3$};
            \draw (0,1) -- (1,1) node[anchor=south]{$\ell_4$};
            
            \draw (2,0) -- (2,1) node[anchor=east]{$n_1$};
            \draw (3,0) -- (3,1) node[anchor=west]{$n_2$};
            \draw (2,0) -- (3,0) node[anchor=north]{$n_3$};
            \draw (2,1) -- (3,1) node[anchor=south west]{$n_4$};
        \end{scope}
    \end{tikzpicture}
        \label{fg:matter:example}
    \end{subfigure}
    \hfill
    \begin{subfigure}[t]{0.18\textwidth}
        \caption{)}
        \begin{tikzpicture}[scale=.65]
        \draw[very thin] (-.5,-.5) grid (2.5,3.5);
        \draw[fill,red] (0,0) circle (.1) node[black,scale=.7,anchor=north east]{$j_1$};
        \draw[fill,red] (2,3) circle (.1) node[black,scale=.6,anchor=south west]{$j_2$};
        \begin{scope}[thick,red,every node/.style={anchor= north,pos=.5,scale=.7,black}]
            \draw (0,0) -- (1,0) node{$\ell_1$};
            \draw (1,0)-- (2,0) node{$\ell_2$};
        \end{scope}
        \begin{scope}[thick,red,every node/.style={anchor= west,pos=.5,scale=.6,black}]
            \draw (2,0) -- (2,1) node{$\ell_3$};
            \draw (2,1) -- (2,2) node{$\ell_4$};
            \draw (2,2) -- (2,3) node{$\ell_5$};
        \end{scope}
        \draw[fill] (1,0) circle (.07);
            \draw[fill] (2,0) circle (.07);
            \draw[fill] (2,1) circle (.07);
            \draw[fill] (2,2) circle (.07);
\end{tikzpicture}
        \label{fg:matter:path}
    \end{subfigure}
    \hfill
    \begin{subfigure}[t]{0.45\textwidth}
        \caption{)}
        \begin{tikzpicture}[scale=0.4]
    \draw[<->, thick] (0.5,2) arc (180:90:2) node[anchor=east,pos=0.45,scale=.9,xshift=+1mm,yshift=+2mm]
        {$\hat{\mathcal{P}}_{\ell_4}\otimes\hat{\mathcal{P}}_{n_4}$};
    \draw[<->, thick] (9.5,2) arc (0:90:2) node[anchor=west,pos=.45,scale=.9,,xshift=-1mm,yshift=+2mm]
        {$\hat{\mathcal{P}}_{\ell_2}\otimes\hat{\mathcal{P}}_{n_2}$};
    \begin{tzket}
        \draw[fill,color=red!20] (0,0) rectangle (1,1);
        \draw[fill,color=blue!20] (2,0) rectangle (3,1);
        \begin{scope}[every node/.style={sloped,allow upside down,scale=.4}]
            \draw (0,0) -- node{\midarrow} (0,1);
            \draw (1,1) -- node{\midarrow} (1,0);
            \draw (2,0) -- node{\midarrow} (2,1);
            \draw (3,1) -- node{\midarrow} (3,0);
            
            \draw (0,0) -- node{\midarrow} (1,0);        
            \draw (0,1) -- node{\midarrow} (1,1);
            
            \draw (2,0) -- node{\midarrow} (1,0);        
            \draw (1,1) -- node{\midarrow} (2,1);
            
            \draw (2,0) -- node{\midarrow} (3,0);        
            \draw (2,1) -- node{\midarrow} (3,1);
        \end{scope}
        \draw[fill, color=red] (2,0) circle (0.15);
        \draw[fill, color=blue] (1,0) circle (0.15);
        \draw[fill, color=blue] (+3,0) circle (0.15);
        \draw (1.5,-.5) node[anchor=north,scale=.8]{\rotatebox{270}{$=$}};
        \draw (1.5,-1.2) node[anchor=north,scale=.8]{$ \ket{\phi'_1}$};
        \tzwidth{3}
    \end{tzket}

    \begin{tzket}[shift={(3.5,4)}]
        \draw[fill,color=red!20] (0,0) rectangle (1,1);
        \draw[fill,color=blue!20] (2,0) rectangle (3,1);
        \begin{scope}[every node/.style={sloped,allow upside down,scale=.4}]
            \draw (0,0) -- node{\midarrow} (0,1);
            \draw (1,1) -- node{\midarrow} (1,0);
            \draw (2,0) -- node{\midarrow} (2,1);
            \draw (3,1) -- node{\midarrow} (3,0);
            
            \draw (0,0) -- node{\midarrow} (1,0);        
            \draw (1,1) -- node{\midarrow} (0,1);
            
            \draw (2,0) -- node{\midarrow} (1,0);        
            \draw (1,1) -- node{\midarrow} (2,1);
            
            \draw (2,0) -- node{\midarrow} (3,0);        
            \draw (3,1) -- node{\midarrow} (2,1);
        \end{scope}
        \draw[fill, color=red] (2,0) circle (0.15);
        \draw[fill, color=red] (1,1) circle (0.15);
        \draw[fill, color=red] (3,1) circle (0.15);
        \draw[fill, color=blue] (1,0) circle (0.15);
        \draw[fill, color=blue] (3,0) circle (0.15);
        \draw[fill, color=blue] (0,1) circle (0.15);    
        \draw[fill, color=blue] (2,1) circle (0.15);
        \draw (1.5,-.5) node[anchor=north,scale=.8]{\rotatebox{270}{$=$}};
        \draw (1.5,-1.2) node[anchor=north,scale=.8]{$ \ket{\Phi'}$};
        \tzwidth{3}
    \end{tzket}
    \begin{tzket}[shift={(7,0)}]
        \draw[fill,color=red!20] (0,0) rectangle (1,1);
        \draw[fill,color=blue!20] (2,0) rectangle (3,1);
        \begin{scope}[every node/.style={sloped,allow upside down,scale=.4}]
            \draw (0,0) -- node{\midarrow} (0,1);
            \draw (1,0) -- node{\midarrow} (1,1);
            \draw (2,0) -- node{\midarrow} (2,1);
            \draw (3,0) -- node{\midarrow} (3,1);
            
            \draw (0,0) -- node{\midarrow} (1,0);        
            \draw (1,1) -- node{\midarrow} (0,1);
            
            \draw (2,0) -- node{\midarrow} (1,0);        
            \draw (1,1) -- node{\midarrow} (2,1);
            
            \draw (2,0) -- node{\midarrow} (3,0);        
            \draw (3,1) -- node{\midarrow} (2,1);
        \end{scope}
        \draw[fill, color=red] (2,0) circle (0.15);
        \draw[fill, color=blue] (0,1) circle (0.15);
        \draw[fill, color=blue] (2,1) circle (0.15);
        \draw (1.5,-.5) node[anchor=north,scale=.8]{\rotatebox{270}{$=$}};
        \draw (1.5,-1.2) node[anchor=north,scale=.8]{$ \ket{\phi'_2}$};
        \tzwidth{3}
    \end{tzket}
\end{tikzpicture}
        \label{fg:matter:entangled}
    \end{subfigure}
    \vfill
    \vspace{-10mm}
    \begin{subfigure}[t]{0.2\textwidth}
        \caption{)}
        \begin{tikzpicture}[scale=.6]  
            \draw[fill,color=red!20] (0,0) rectangle (1,1);
            \draw[fill,color=blue!20] (2,0) rectangle (3,1);
            \draw[very thin] (-.5,-.5) grid (3.5,1.5);
            \foreach \i in {0,1}
            \foreach \j in {0,1,2,3}
            \draw[fill] (\j,\i) circle (.05);
            \draw (0,0) node[anchor = north east,scale=.6]{$-1$};
            \draw (1,0) node[anchor = north east,scale=.6]{$0$};
            \draw (1,1) node[anchor = south west,scale=.6]{$-1$};
            \draw (0,1) node[anchor = south west,scale=.6]{$0$};
        
            \draw (2,0) node[anchor = north east,scale=.6]{$0$};
            \draw (3,0) node[anchor = north east,scale=.6]{$0$};
            \draw (3,1) node[anchor = south west,scale=.6]{$0$};
            \draw (2,1) node[anchor = south west,scale=.6]{$+1$};
            \draw (3.5,.5) node[anchor=west,scale=.8]{$\xrightarrow{\mathcal{H}_{\mathcal{AB},1}}$};
\end{tikzpicture}

\begin{tikzpicture}[scale=.6]  
         \draw[fill,color=red!20] (0,0) rectangle (1,1);
            \draw[fill,color=blue!20] (2,0) rectangle (3,1);
            \draw[very thin] (-.5,-.5) grid (3.5,1.5);
            \foreach \i in {0,1}
            \foreach \j in {0,1,2,3}
            \draw[fill] (\j,\i) circle (.05);
       
            \draw (0,0) node[anchor = north east,scale=.6]{$-1$};
            \draw (1,0) node[anchor = north east,scale=.6]{$+1$};
            \draw (1,1) node[anchor = south west,scale=.6]{$-1$};
            \draw (0,1) node[anchor = south west,scale=.6]{$0$};
        
            \draw (2,0) node[anchor = north east,scale=.6]{$-1$};
            \draw (3,0) node[anchor = north east,scale=.6]{$0$};
            \draw (3,1) node[anchor = south west,scale=.6]{$0$};
            \draw (2,1) node[anchor = south west,scale=.6]{$+1$};
        \draw (3.5,.5) node[anchor=west,scale=.8]{$\xrightarrow{\mathcal{H}_{\mathcal{AB},2}}$};
\end{tikzpicture}

\begin{tikzpicture}[scale=.6]  
         \draw[fill,color=red!20] (0,0) rectangle (1,1);
            \draw[fill,color=blue!20] (2,0) rectangle (3,1);
            \draw[very thin] (-.5,-.5) grid (3.5,1.5);
            \foreach \i in {0,1}
            \foreach \j in {0,1,2,3}
            \draw[fill] (\j,\i) circle (.05);
            
            \draw (0,0) node[anchor = north east,scale=.6]{$-1$};
            \draw (1,0) node[anchor = north east,scale=.6]{$0$};
            \draw (1,1) node[anchor = south west,scale=.6]{$0$};
            \draw (0,1) node[anchor = south west,scale=.6]{$0$};
        
            \draw (2,0) node[anchor = north east,scale=.6]{$0$};
            \draw (3,0) node[anchor = north east,scale=.6]{$0$};
            \draw (3,1) node[anchor = south west,scale=.6]{$0$};
            \draw (2,1) node[anchor = south west,scale=.6]{$0$};
      
        \draw (3.5,.5) node[anchor=west,scale=.8]{$\xrightarrow{\mathcal{H}_{\mathcal{AB},3}}$};
\end{tikzpicture}

\begin{tikzpicture}[scale=.6]  
         \draw[fill,color=red!20] (0,0) rectangle (1,1);
            \draw[fill,color=blue!20] (2,0) rectangle (3,1);
            \draw[very thin] (-.5,-.5) grid (3.5,1.5);
            \foreach \i in {0,1}
            \foreach \j in {0,1,2,3}
            \draw[fill] (\j,\i) circle (.05);
            
            \draw (0,0) node[anchor = north east,scale=.6]{$-1$};
            \draw (1,0) node[anchor = north east,scale=.6]{$+1$};
            \draw (1,1) node[anchor = south west,scale=.6]{$0$};
            \draw (0,1) node[anchor = south west,scale=.6]{$0$};
        
            \draw (2,0) node[anchor = north east,scale=.6]{$-1$};
            \draw (3,0) node[anchor = north east,scale=.6]{$0$};
            \draw (3,1) node[anchor = south west,scale=.6]{$0$};
            \draw (2,1) node[anchor = south west,scale=.6]{$0$};
      
        \draw (3.5,.5) node[anchor=west,scale=.8]{$\xrightarrow{\mathcal{H}_{\mathcal{AB},4}}$};
\end{tikzpicture}
        \label{fg:matter:sectors}
    \end{subfigure}
    \hfill
    \begin{subfigure}[t]{0.75\textwidth}
        \caption{)}
        \input{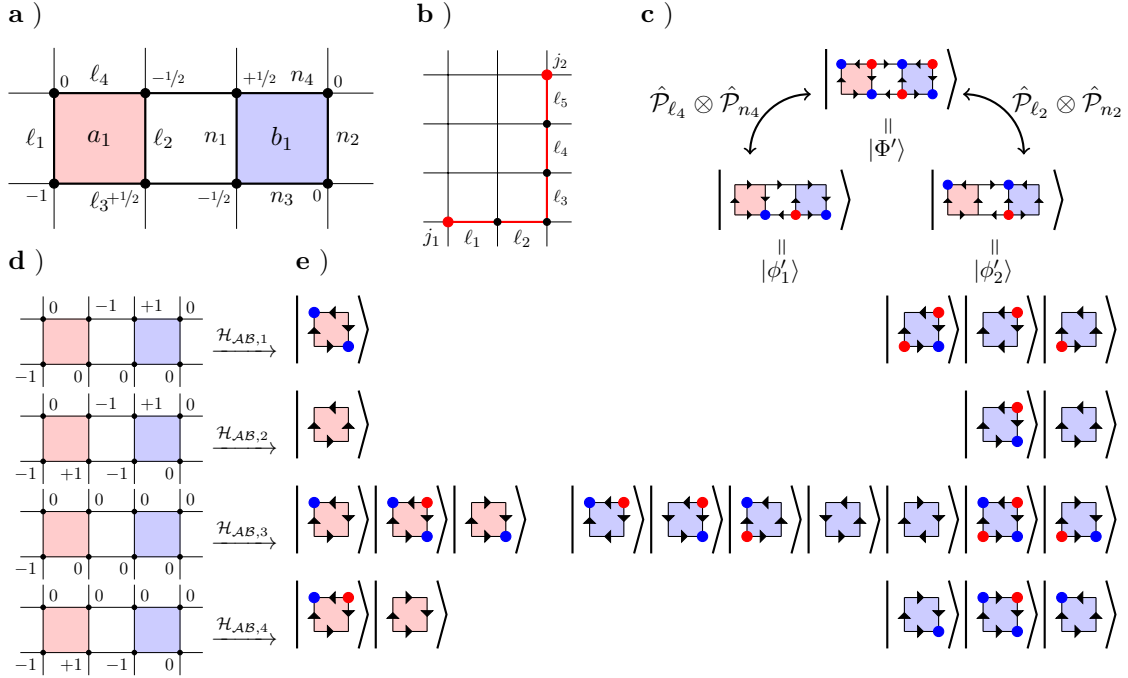}
        \label{fg:matter:bases}
    \end{subfigure}
    \caption{\textbf{(a)} Geometry for \(\lat{L}\), SS, and sub-regions \(\lat{A}\) and \(\lat{B}\) chosen in the example calculations in presence of dynamical fermions.
    \textbf{(b)} Example of a path, $\Gamma = [\ell_1,\dots,\ell_5]$, connecting two lattice points $j_1$ and $j_2$.
    Dynamical fermionic matter lives on the lattice sites, while gauge fields live on links between sites.
    \textbf{(c)} A linear combination of the three states depicted yields an entangled state of a \(\mathrm{U}(1)\) gauge theory in presence of dynamical fermionic matter.
    The entanglement can be witnessed by the operator \(\hat{C}'\) defined in equation~\eqref{eq:matter:witness}.
    Its constituents are path operators \(\pathob{\ell_4}\otimes\pathob{n_4}\) and \(\pathob{\ell_2}\otimes\pathob{n_2}\) (see equation~\eqref{eq:matter:path}), which connect the three states as illustrated.
    \textbf{(d)} Compatible pairs of SSs, \(\hsss[,n]{A}\otimes \hsss[,n]{B}\) with \(n=1,\dots,4\), associated with regions \(\lat{A}\) and \(\lat{B}\).
    \textbf{(e)} Choice of basis for the SSs in \ref{fg:matter:sectors}.
    States  are illustrations of \(\ket{\varphi_{\lat{X},n}^j}\) and span $\hsss[,n]{X}$.
    Negative (positive) charges can only live on odd (even) sites and whenever they are present $\hat{\psi}^\dag\hat{\psi}$ takes the value 0 (1), marked by a red (blue) bullet.}
    \label{fg:matter}
\end{figure}

The electric field and plaquette operators defined in equations~\eqref{eq:gauge:generator} and~\eqref{eq:gauge:plaquette}, respectively, remain invariant under the action of the gauge generators in equation~\eqref{eq:matter:generator}.
In addition, the presence of the dynamic fermions allows for new gauge invariant operators.
Besides local particle densities $\hat{\psi}^\dag_j \hat{\psi}_j$ (and products thereof), these are fermionic correlators connected by gauge strings of the form 
\begin{align}
    \pathop{\Gamma} = \hat{\psi}^\dag_{j_1} (\prod_{m = 1}^n
    \hat{\sigma}^\dag_{\ell_m}) \hat{\psi}_{j_2} 
    \label{eq:matter:path}\,,
\end{align} 
where $\Gamma = [\ell_1,\dots,\ell_n]$ denotes a path consisting of $n$ links connecting lattice points $j_1$ and $j_2$, such as illustrated in figure~\ref{fg:matter:path}.
Such operators commute with every $\hat{g}_j$ and thus are gauge invariant.
Importantly, both endpoints of such fermionic correlators need to lie within a contiguous region, unlike, e.g., for fermionic systems without gauge symmetry, where one can choose correlators between arbitrary modes, which can also lie in disjoint spatial regions.

In analogy with section~\ref{sec:gauge:bipartition}, given a region $\lat{L}$ and one of its boundary points $j$, we define elements of $\galg{L}$ as
\begin{equation}
    \hat{Z}_j = \hat{\psi}^\dag_j\hat{\psi}_j - \frac{1-(-1)^{(x_j+y_j)}}{2} - 
    \sum_{\ell \in \lat{L}}\mathrm{sign}(\ell,j)\hat{E}_\ell\,.
    \label{eq:matter:center}
\end{equation}
Figure~\ref{fg:matter:example} exemplifies a region \(\lat{L}\) and choice of SS \(\hsss{L}\), as well as two disjoint sub-regions \(\lat{A}\) and \(\lat{B}\) .
The corresponding pairs of compatible SSs \((\sss{A},\sss{B})\) are represented in figure~\ref{fg:matter:sectors}, while figure~\ref{fg:matter:bases} sketches the basis we adopt to perform explicit calculations.

\subsection{Construction of an entanglement witness}
\label{sec:matter:witness}

With these definitions, we can construct entanglement witnesses for a gauge theory in presence of dynamical matter.
The operator \(\hat{C}=\plaqob{a_1} \otimes \plaqob{b_1}\), defined as in equation~\eqref{eq:gauge:witness}, is still gauge-invariant and thus remains valid in principle as a entanglement witness.
However, its action on the examples in figure~\ref{fg:matter:entangled} vanishes for the chosen SS of \(\lat{L}\).
Hence, it is not useful to witness entanglement.

We modify the witness to include path operators \(\pathob{\ell} = \pathop{\ell} +  \pathop{\ell}^\dag\) for neighboring lattice sites separated by a bond \(\ell\) (see equation~\eqref{eq:matter:path}).
One could also consider more general definitions, such as \(\pathob{\ell} = \pathop{\ell} e^{i\phi_\ell} + \pathop{\ell}^\dag e^{-i\phi_\ell}\).
Nonetheless, for the sake of simplicity and with no loss of generality, we fix $\phi_\ell=0$ for all bonds.
Labeling the links as in figure~\ref{fg:matter:example}, we define 
\begin{equation}
    \hat{C}' = \plaqob{a_1}\otimes\plaqob{b_1} + \sum_{i = 1}^4 \pathob{\ell_i} \otimes \pathob{n_i}\,.
    \label{eq:matter:witness}
\end{equation}
The extremal eigenvalues of \(\hat{C}'\) are \(\omega^- = -\sqrt{2}\) and \(\omega^+=+\sqrt{2}\).
To determine bounds of \(\hat{C}'\) on a separable state in $\hsss[,n]{A}\otimes\hsss[,n]{B}$ (i.e., $\omega_\mathrm{sep}^-{(\sss[,n]{A},\sss[,n]{B})}$ and $\omega_\mathrm{sep}^+(Z^\mathcal{A}_n,Z^\mathcal{B}_n)$), we consider each case of figure~\ref{fg:matter:sectors} separately.
To this end, we evaluate \(f(\bm{a},\bm{b},\bm{\alpha},\bm{\beta})=\bra{\Phi_n}\hat{C}'\ket{\Phi_n}\) on a product state \(\ket{\Phi_n} = \ket{\phi_{\lat{A},n}} \otimes\ket{\phi_{\lat{B},n}}\) with \(\ket{\phi_{\lat{A},n}} = \sum_j a_j e^{i\alpha_j} \ket{\varphi_{\lat{A},n}^j}\) and \(\ket{\phi_{\lat{A},n}} = \sum_j b_j e^{i\beta_j}\ket{\varphi_{\lat{B},n}^j}\).
We optimize \(f(\cdot)\) under the constraints given by the normalization of \(\ket{\phi_{\lat{A},n}}\) and \(\ket{\phi_{\lat{B},n}}\) (i.e., \(\sum_j a_j^2 =1\) and \(\sum_j b_j^2 = 1\)).
For small systems, the calculations can be performed analytically, while for larger regions we use an optimization approach detailed in appendix~\ref{sec:appendix_C}.
The explicit results for the simple case depicted in figure~\ref{fg:matter:example} are as follows.

\paragraph{First and second pair of SSs (see figure~\ref{fg:matter:sectors})}
The action of \(\hat{C}'\) vanishes on both pairs \(\hsss[,1]{A}\otimes\hsss[,1]{B}\) and \(\hsss[,2]{A}\otimes\hsss[,2]{B}\), so \(f(\bm{a},\bm{b},\bm{\alpha},\bm{\beta})= 0\).
It follows that \(\omega_\mathrm{sep}^\pm(\sss[,1]{A},\sss[,1]{B}) = \omega_\mathrm{sep}^\pm(\sss[,2]{A},\sss[,2]{B}) = 0\).
\paragraph{Third pair of SSs (see figure~\ref{fg:matter:sectors})}
Defining $\alpha_{ij} \,=\, \alpha_i - \alpha_j$ and $\beta_{ij}\, = \beta_i - \beta_j$, we have
\begin{align}
    f(\bm{a},\bm{b},\bm{\alpha},\bm{\beta}) &= 4 a_2 a_3 \cos\alpha_{23} [b_2b_4 \cos\beta_{24} + b_3b_6\cos\beta_{36}]\nonumber\\
    &\quad+ 4 a_1 a_2 \cos\alpha_{23} [b_1b_5\cos\beta_{15}+ b_6b_7\cos\beta_{67}]
    \,,
\end{align}
for the action of \(\hat{C}'\) on \(\hsss[,3]{A}\otimes\hsss[,3]{B}\).
Since the cosine takes values in \([-1,+1]\), we obtain \(|f(\bm{a},\bm{b},\bm{\alpha},\bm{\beta})| \leq 4 a_2 a_3 (b_2 b_4 + b_3 b_6) + 4 a_1 a_2 (b_1 b_5+ b_6 b_7)\leq 1 \).
Consequently, the bounds are \(\omega_\mathrm{sep}^\pm(\sss[,3]{A},\sss[,3]{B}) = \pm1\).
\paragraph{Fourth pair of SSs (see figure~\ref{fg:matter:sectors}).}
We have \(f(\bm{a},\bm{b},\bm{\alpha},\bm{\beta}) = 4 a_1 a_2 \cos \alpha_{12} \cdot b_1 b_2 \cos\beta_{12}\) for \(\hsss[,4]{A}\otimes\hsss[,4]{B}\), and thus \(|f(\bm{a},\bm{b},\bm{\alpha},\bm{\beta})|\leq 4 a_1a_2b_1b_2 \leq 1\).
Therefore, the bound is also give by \(\omega_\mathrm{sep}^\pm(\sss[,4]{A},\sss[,4]{B}) = \pm1\).

In conclusion, optimizing over all SSs, we find \(\omega_\mathrm{sep}^\pm = \pm 1\).
Thus, any separable state between regions \(\lat{A}\) and \(\lat{B}\) must satisfy \(-1\leq\langle\hat{C}'\rangle_\mathrm{sep} \leq 1\).
With these inequalities, we can define the entanglement witnesses \(\hat{W}'_\pm = \pm( \omega_\mathrm{sep}^\pm \cdot \hat{\mathbb{1}} - \hat{C}')\).

\subsection{Example of correctly detected entangled state}
As an example of a state that is detected as entangled by $\hat{W}'_+$, we consider  $\ket{\Psi} = \frac 12 \ket{\phi'_1} + \frac{1}{\sqrt{2}}\ket{\Phi'} +\frac 12 \ket{\phi'_2}$, as illustrated in figure~\ref{fg:matter:entangled}.
All three states have an excess positive charge in region $\mathcal{A}$ and are charge neutral in $\mathcal{B}$.
From inspection of figure~\ref{fg:matter:entangled}, one sees that $\ket{\Psi}$ is non-separable.
We have that \(\hat{C}'\ket{\Psi} = (\pathob{\ell_2} \otimes\pathob{n_2} + \pathob{\ell_4}\otimes \pathob{n_4})\ket{\Psi} = \sqrt{2}\ket{\Psi}\).
Thus, the expectation value of \(\hat{W}_+\) on \(\ket{\Psi}\) is \(1-\sqrt{2} < 0\), and \(\ket{\Psi}\) is correctly recognized as entangled.

\section{Discussion}
\label{sec:discussion}

In this work, we have introduced the concept of entanglement witnessing for lattice gauge theories.
A key challenge derives from the absence of a tensor product structure of the Hilbert space, which is rather given by the direct sum of different gauge SSs.
This issue makes it impractical to define separability directly at the state level.
To resolve this challenge, we have adopted a definition within which separable regions are those where all correlators between gauge-invariant operators decompose into products. Turning the typical approach of tensor-product Hilbert spaces around, this requirement of factorization then defines the structure of separable states, and it leads us to a natural generic form of entanglement witnesses for LGTs. 

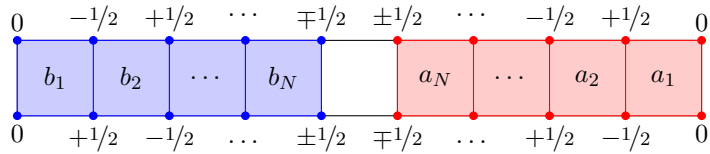
\begin{figure}
    \centering
    \begin{tikzpicture}[scale=1]  
    \draw (0,0) grid (9,1);
    \draw[fill, color=blue!20] (0,0) rectangle(4,1);
    \draw[blue] (0,0) grid (4,1);
    \draw[fill, color=red!20] (5,0) rectangle (9,1);
    \draw[red] (5,0) grid (9,1);
    
    \begin{scope}[every node/.style={scale=.9,anchor=north}]
        \draw (0,0) node{$0$};
        \draw (1,0) node{$+\nicefrac{1}{2}$};
        \draw (2,0) node{$-\nicefrac{1}{2}$};
        \draw (3,-.2) node{\dots};
        \draw (4,0) node{$\pm\nicefrac{1}{2}$};
        \draw (5,0) node{$\mp\nicefrac{1}{2}$};
        \draw (6,-.2) node{\dots};
        \draw (7,0) node{$+\nicefrac{1}{2}$};
        \draw (8,0) node{$-\nicefrac{1}{2}$};
        \draw (9,0) node{$0$};
    \end{scope}
    \foreach \i in {0,...,4} \foreach \j in {0,1}
        \draw[fill, blue] (\i,\j) circle (.05);
        
    \foreach \i in {5,...,9} \foreach \j in {0,1}
        \draw[fill, red] (\i,\j) circle (.05);
    
    \begin{scope}[every node/.style={scale=.9,anchor=south}]
        \draw (0,1) node{$0$};
        \draw (1,1) node{$-\nicefrac{1}{2}$};
        \draw (2,1) node{$+\nicefrac{1}{2}$};
        \draw (3,1.2) node{\dots};
        \draw (4,1) node{$\mp\nicefrac{1}{2}$};
        \draw (5,1) node{$\pm\nicefrac{1}{2}$};
        \draw (6,1.2) node{$\dots$};
        \draw (7,1) node{$-\nicefrac{1}{2}$};
        \draw (8,1) node{$+\nicefrac{1}{2}$};
        \draw (9,1) node{0};
    \end{scope}
    \foreach \i in {1,2}
        \draw (\i-.5,.5) node[scale=.9]{$b_\i$};
    \draw (2.5,.5) node[scale=.9]{\dots};
    \draw (3.5,.5) node[scale=.9]{$b_N$};
    
    \foreach \i in {1,2}
        \draw (9.5 -\i,.5) node[scale=.9]{$a_\i$};
    \draw (6.5,.5) node[scale=.9]{\dots};
    \draw (5.5,.5) node[scale=.9]{$a_N$};
\end{tikzpicture}
    \caption{
        Geometries and SS choices considered for the resource scaling summarized in table~\ref{tb:resource}.
        The total number of plaquettes is \(N_\mathrm{plaq} = 2 N +1\), with \(N\) in subregions \(\lat{A}\) and \(\lat{B}\) each.
    }
    \label{fg:resource}
\end{figure}

One central advantage of our framework is that the evaluation of the witnesses considered here requires only polynomial resources --- as is typical for entanglement witnesses \cite{friis_vitagliano_malik_huber_2018,guhne_toth_2009}.
This is in sharp contrast to the exponential scaling of full-fledged entanglement measures. 
For instance, it is instructive to compare the effort required to carry out full tomography with that of entanglement witnessing using the operator in equation~\eqref{eq:gauge:witness}.
The results for the geometries and border conditions sketched in figure~\ref{fg:resource} are summarized in table~\ref{tb:resource}.

\begin{table}
    \small
    \centering
    \begin{tabular}{|m{0.1\linewidth} m{0.1\linewidth} m{0.1\linewidth} m{0.1\linewidth} m{0.1\linewidth} m{0.1\linewidth}|}
        \hline
        \(N\)& 2 & 3 & 4 & 5 & 6\\
        \(N_\mathrm{plaq}\)& 5 & 7 & 9 & 11 & 13\\
        \hline
        \(N_\mathrm{ft}\) & 168 & 1155 & 7920 & 54288 & 372099\\
        \(N_\mathrm{pt}\) & 8 & 24 & 63 & 168 & 440\\
        \(N_\mathrm{ew}\) & 2 & 3 & 4 & 5 & 6\\\hline
    \end{tabular}
    \caption[justification=justified]{
    Resource scaling of entanglement certification for the systems sketched in figure~\ref{fg:resource}.
    The number of measurements required for full tomography of the system \(\lat{L}\), with Hilbert space dimension \(d_{\lat{L}}\), grows exponentially with system size according to \(N_\mathrm{ft}=d_\lat{L}^2-1\).
    Full-fledged entanglement measures on a subsystem \(\lat{A}\) can be computed by expressing the reduced density matrix of \(\lat{A}\) as a mixture of SSs \(\hsss{A}\).
    However, partial tomography of the reduced density matrix of subsystem \(\lat{A}\) is also typically plagued by exponentially growing cost (see, e.g., \cite{zache_van_damme_halimeh_hauke_banerjee_2022} for an example in 1+1D).
    Specifically, the number of measurements necessary is given by \(N_\mathrm{pt}=d_\lat{A}^2-1\) where \(d_\lat{A}\) is the dimension of the chosen SS \(\hsss{A}\) (illustrated here for the largest superselection sector).
    In contrast, the number of independent measurements to evaluate the entanglement witness \(\hat{C}\) defined in  equation~\eqref{eq:gauge:witness} only scales polynomially as \(N_\mathrm{ew}= (N_\mathrm{plaq} -1 )/2\).
    Depending on the specific scenario, this number can be further decreased, e.g., by parallel measurements of commuting summands of \(\hat{C}\).
    As these figures illustrate, entanglement witnesses for LGTs can drastically reduce the resource cost of entanglement detection.
    }
    \label{tb:resource}
\end{table}

Though we have illustrated concrete constructions for \(\mathrm{U}(1)\) gauge symmetry in two spatial dimensions, our concept is general.
As such, there are many possibilities for future theoretical developments.
In particular, the presented scheme can be applied to other symmetries, such as discrete groups. An especially interesting set of questions to which the framework may contribute evolves around the presence of entanglement in non--Abelian models, for which it has been found that entanglement entropy and entanglement distillation behave qualitatively different from Abelian models~\cite{ghosh_soni_trivedi_2015,Soni2016JHEP,van_acoleyen_bultinck_haegeman_marien_scholz_verstraete_2016}.
Another potential setting for applying our framework is in topologically ordered models where the gauge symmetry is enforced dynamically. 

Finally, our work opens the way to certifying entanglement in a fully scalable way in modern quantum simulators of LGTs~\cite{alexeev_bacon_brown_calderbank_carr_chong_demarco_englund_farhi_fefferman_et_al_2021,aidelsburger_barbiero_bermudez_chanda_dauphin_et_al_2021,zohar_2021,banuls_blatt_catani_celi_cirac_dalmonte_fallani_jansen_lewenstein_montangero_et_al_2020,mil_zache_hegde_xia_bhatt_oberthaler_hauke_berges_jendrzejewski_2020,mildenberger_mruczkiewicz_halimeh_jiang_hauke_2022,nguyen_tran_zhu_green_alderete_davoudi_linke_2022}.
An interesting application will be, e.g., to study the role of entanglement build-up in equilibration dynamics of gauge theories~\cite{zhou_su_halimeh_ott_sun_hauke_yang_yuan_berges_pan_2022,berges_floerchinger_venugopalan_2018}, which is a major open question in contemporary heavy--ion collision experiments~\cite{feal_pajares_vazquez_2019,bellwied_2018}. 

\paragraph{Acknowledgements.---} We acknowledge support by the ERC Starting Grant StrEnQTh (project ID 804305), Provincia Autonoma di Trento, and by Q@TN, the joint lab between University of Trento, FBK-Fondazione Bruno Kessler, INFN-National Institute for Nuclear Physics and CNR-National Research Council.

\bibliographystyle{ieeetr}
\bibliography{main}

\appendix
\section{Conditions for a SS to be physical} 
\label{sec:appendix_A}
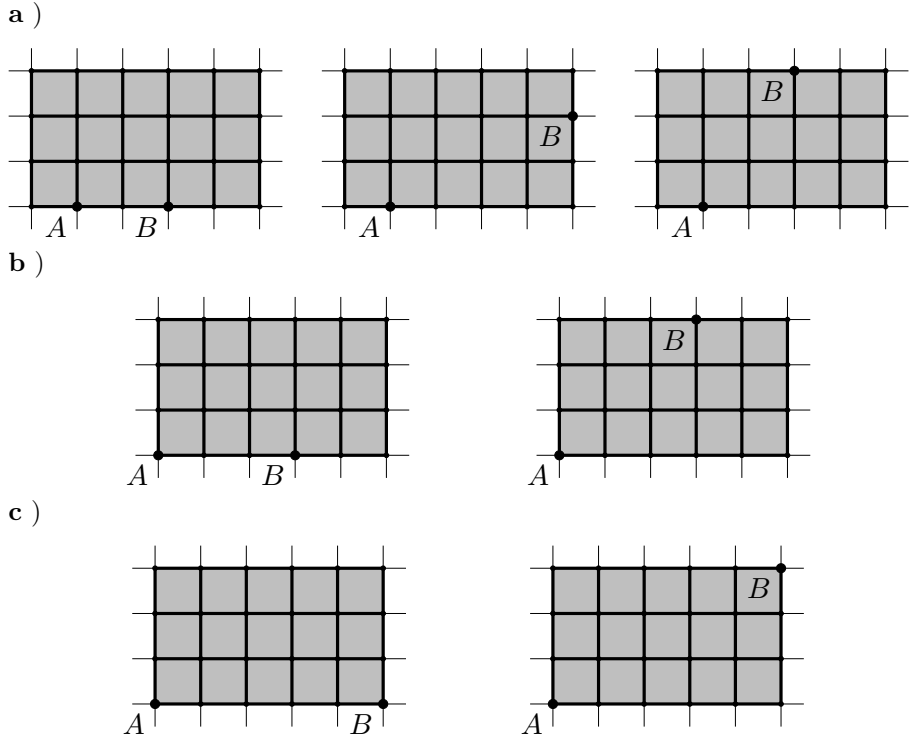
\begin{figure}[h]
    \centering
    \begin{subfigure}[t]{.8\textwidth}
        \caption{)}
        \begin{tikzpicture}[scale=.6]
    \draw[very thin] (-.5,-.5) grid (5.5,3.5);
    \draw[fill, color=gray!50] (0,0) rectangle (5,3);
    \draw[very thick] (0,0) grid (5,3);
    \foreach \i in {0,...,5}
    \foreach \j in {0,...,3}
    \draw[fill] (\i,\j) circle (.05);
    \draw[fill] (1,0) circle (.1) node[anchor=north east]{$A$};
    \draw[fill] (3,0) circle (.1) node[anchor=north east]{$B$};
\end{tikzpicture}\hfill
\begin{tikzpicture}[scale=.6]
    \draw[very thin] (-.5,-.5) grid (5.5,3.5);
    \draw[fill, color=gray!50] (0,0) rectangle (5,3);
    \draw[very thick] (0,0) grid (5,3);
    \foreach \i in {0,...,5}
    \foreach \j in {0,...,3}
    \draw[fill] (\i,\j) circle (.05);
    \draw[fill] (1,0) circle (.1) node[anchor=north east]{$A$};
    \draw[fill] (5,2) circle (.1) node[anchor=north east]{$B$};
\end{tikzpicture}\hfill
\begin{tikzpicture}[scale=.6]
\draw[very thin] (-.5,-.5) grid (5.5,3.5);
\draw[fill, color=gray!50] (0,0) rectangle (5,3);
\draw[very thick] (0,0) grid (5,3);
\foreach \i in {0,...,5}
\foreach \j in {0,...,3}
\draw[fill] (\i,\j) circle (.05);
\draw[fill] (1,0) circle (.1) node[anchor=north east]{$A$};
\draw[fill] (3,3) circle (.1) node[anchor=north east]{$B$};
\end{tikzpicture}  
        \label{fg:A:a}
    \end{subfigure}
    
    \begin{subfigure}[t]{.8\textwidth}
        \caption{)}
        \hfill\begin{tikzpicture}[scale=.6]
    \draw[very thin] (-.5,-.5) grid (5.5,3.5);
    \draw[fill, color=gray!50] (0,0) rectangle (5,3);
    \draw[very thick] (0,0) grid (5,3);
    \foreach \i in {0,...,5}
    \foreach \j in {0,...,3}
    \draw[fill] (\i,\j) circle (.05);
    \draw[fill] (0,0) circle (.1) node[anchor=north east]{$A$};
    \draw[fill] (3,0) circle (.1) node[anchor=north east]{$B$};
\end{tikzpicture}\hfill
\begin{tikzpicture}[scale=.6]
    \draw[very thin] (-.5,-.5) grid (5.5,3.5);
    \draw[fill, color=gray!50] (0,0) rectangle (5,3);
    \draw[very thick] (0,0) grid (5,3);
    \foreach \i in {0,...,5}
    \foreach \j in {0,...,3}
    \draw[fill] (\i,\j) circle (.05);
    \draw[fill] (0,0) circle (.1) node[anchor=north east]{$A$};
    \draw[fill] (3,3) circle (.1) node[anchor=north east]{$B$};
\end{tikzpicture}\hfill
        \label{fg:A:b}
    \end{subfigure}
    
    \begin{subfigure}[t]{.8\textwidth}
        \caption{)}
        \hfill
\begin{tikzpicture}[scale=.6]
    \draw[very thin] (-.5,-.5) grid (5.5,3.5);
    \draw[fill, color=gray!50] (0,0) rectangle (5,3);
    \draw[very thick] (0,0) grid (5,3);
    \foreach \i in {0,...,5}
    \foreach \j in {0,...,3}
    \draw[fill] (\i,\j) circle (.05);
    \draw[fill] (0,0) circle (.1) node[anchor=north east]{$A$};
    \draw[fill] (5,0) circle (.1) node[anchor=north east]{$B$};
\end{tikzpicture}\hfill
\begin{tikzpicture}[scale=.6]
    \draw[very thin] (-.5,-.5) grid (5.5,3.5);
    \draw[fill, color=gray!50] (0,0) rectangle (5,3);
    \draw[very thick] (0,0) grid (5,3);
    \foreach \i in {0,...,5}
    \foreach \j in {0,...,3}
    \draw[fill] (\i,\j) circle (.05);
    \draw[fill] (0,0) circle (.1) node[anchor=north east]{$A$};
    \draw[fill] (5,3) circle (.1) node[anchor=north east]{$B$};
\end{tikzpicture}
\hfill
        \label{fg:A:c}
    \end{subfigure}
    
    \caption{Relative position of points A and B. $\textbf{a)}A \wedge B$ not corner points (left: same edge, middle: adjacent edges, top: opposite edges). \textbf{b)} $A \dot{\vee} B$ is corner point (left: same edge, right: not on same edge). Two generic situations need to be distinguished. \textbf{c)} $A \wedge B$ are corner points (left: same edge, right: not on same edge).}
\end{figure}

Given a rectangular region $\mathcal{R}$, we consider a pair of boundary points $A,B \in \partial_0\mathcal{R}$. We define
\begin{align}
    &\hat{s}_\mathrm{long}(A,B) = \sum_{j \in \Gamma_\mathrm{long}(A,B)} \hat{Z}_j\,,\\
    &\hat{s}_\mathrm{short}(A,B) =  \sum_{j \in \Gamma_\mathrm{short}(A,B)}\hat{Z}_j\,,
\end{align}
where $\Gamma_\mathrm{short}(A,B), \Gamma_{\mathrm{long}}(A,B) \subset \partial_0\mathrm{R}$ are the paths along the contour of $\mathcal{R}$ that connect $A$ and $B$. Considering a SS labeled by an array $Z$ (i.e., $\mathcal{H}_{\mathcal{R}}(Z)$), we have that, $\forall\ket{\Phi} \in \mathcal{H}_{\mathcal{R}}(Z)$,
\begin{align}
    &\langle\hat{s}_\mathrm{long}(A,B)\rangle_{\Phi} = \sum_{j \in \Gamma_\mathrm{short}(A,B)} Z_j \,,\\
    &\langle\hat{s}_\mathrm{short}(A,B)\rangle_{\Phi} = \sum_{j \in \Gamma_\mathrm{short}(A,B)} Z_j\,.
\end{align}
It is possible to prove that any SS needs to satisfy, for every pair of boundary points, 
the conditions on $\langle\hat{s}_{\Gamma_\mathrm{long}}(A,B)\rangle$ and $\langle\hat{s}_{\Gamma_\mathrm{short}}(A,B)\rangle$ reported in table~\ref{tab:constraints_expression}, where we use the abbreviations $\Delta x = |A_x - B_x|$ and $\Delta y = |A_y - B_y|$, with $A = (A_x, B_y)$ and $B = (B_x,B_y)$.
The conditions depend on the relative position of the boundary points $A$ and $B$, as illustrated in figure~\ref{fg:A:a}, \ref{fg:A:b}, and \ref{fg:A:c}.

\renewcommand{\arraystretch}{1.5}
\begin{table}[h]
\centering
\small
\begin{tabular}{|m{.24\textwidth} m{.3\textwidth} m{.17\textwidth} m{.17\textwidth}|}
    \hline
    \textbf{Position of A and B} &
    \textbf{Specification of position}&
    $\langle\hat{s}_{short}(A,B)\rangle$
    & $\langle\hat{s}_{long}(A,B)\rangle$\\\hline
    \multirow{3}{.24\textwidth}{$A \wedge B$ not on corners (see figure~\ref{fg:A:a})} & same edge &$\pm \frac{\Delta x +\Delta y +1}{2}$ & $\pm \frac{\Delta x + \Delta y - 1}{2}$ \\
    & adjacent edges & $\pm \frac{\Delta x + \Delta y +2}{2}$ & $\pm \frac{\Delta x + \Delta y}{2}$  \\
    &opposite edges  & $\pm \frac{\Delta x + \Delta y + 1}{2}$ & $\pm \frac{\Delta x + \Delta y + 1}{2}$ \\
    \hline
    \multirow{2}{.24\textwidth}{$A \dot{\vee} B$ on corner (see figure~\ref{fg:A:b})}& same edge & $\pm\frac{\Delta x + \Delta y + 2}{2}$ & $\pm\frac{\Delta x + \Delta y - 1}{2}$\\
    & not on same edge &$\pm \frac{\Delta x + \Delta y +1}{2}$ & $\pm \frac{\Delta x + \Delta y } {2}$ \\
    \hline
    \multirow{2}{.24\textwidth}{$A \wedge B$ on corners (see figure~\ref{fg:A:c})} & same edge & $\pm \frac{\Delta x + \Delta y +1}{2}$ & $\pm \frac{\Delta x + \Delta y -1 }{2}$\\
     & not on same edge & $\pm\frac{\Delta x + \Delta y}{2}$&  $\pm\frac{\Delta x + \Delta y}{2}$\\
     \hline
\end{tabular}
\caption{Constraints that an array $Z$ has to verify to represent a physical SS.}
\label{tab:constraints_expression}
\end{table}
\renewcommand{\arraystretch}{1}

\section{Reduction scheme}
\label{sec:appendix_B}

Let $\mathcal{A}$ and $\mathcal{B}$ be two regions with border conditions set by $Z^\mathcal{A}$ and $Z^\mathcal{B}$, respectively. Depending on the SS, the evaluation of the witness defined in equation~\eqref{eq:gauge:witness} can be substantially simplified. 
Suppose that the operator $\hat{\mathcal{U}}_{a_i}$ ($\hat{\mathcal{U}}_{b_i}$) identically vanishes when represented on $\mathcal{H}_\mathcal{A}(Z^\mathcal{A})$ $(\mathcal{H}_\mathcal{B}(Z^\mathcal{B}))$ for $i\in I_\mathcal{A}$ 
($i \in I_\mathcal{B})$, meaning that plaquette $a_i$ ($b_i$) is unflippable.
We remove the pairs of plaquettes $a_i, b_i$ that contain at least one unflippable plaquette from $\mathcal{A}$ and $\mathcal{B}$, obtaining $\mathcal{A}'$ and $\mathcal{B}'$. 
Let $J = [1,\dots,N]  \setminus (I_\mathcal{A}\cup I_\mathcal{B})$, we define $\mathcal{A}' = \bigcup_{j\in J} a_j$ and 
$\mathcal{B}' = \bigcup_{j\in J} b_j$, with new border conditions set through arrays $Z^{\mathcal{A}'}$ and $Z^{\mathcal{B}'}$. 
Then, we define $\mathcal{H}_\mathrm{final}\,\dot{=}\, \mathcal{H}_{\mathcal{A}'}(Z^{\mathcal{A}'})\otimes\mathcal{H}_{\mathcal{B}'}(Z^{\mathcal{B}'})$ and  
$\hat{C}' = \sum_{j\in J} \hat{\mathcal{U}}_{a_j} \otimes \hat{\mathcal{U}}_{b_j}$. 
The initial operator $\hat{C}$ can be written as $\hat{C}' + \sum_{i \in I_\mathcal{A}\cup I_\mathcal{B}}\hat{\mathcal{U}}_{a_i}\otimes\hat{\mathcal{U}}_{b_i}$, where the second term vanishes by hypothesis on $\mathcal{H}_\mathrm{initial}  = \mathcal{H}_\mathcal{A}(Z^\mathcal{A})\otimes\mathcal{H}_\mathcal{B}(Z^\mathcal{B})$. 
Thus, considering a generic state $\hat{\rho}$ in the initial space, we have $\langle\hat{C}\rangle_\rho = Tr_{\mathcal{AB}}(\hat{\rho}\hat{C}) = Tr_{\mathcal{AB}}(\hat{\rho}\hat{C}')$.
To evaluate the expectation value of $\hat{C}$, we can trace out the degrees of freedom not involved in the operator $\hat{C}'$ (i.e., degrees of freedom lying in $\mathcal{A}\setminus \mathcal{A}'$ and in $\mathcal{B}\setminus\mathcal{B}'$), which yields $ \langle\hat{C}\rangle_\rho = Tr_{\mathcal{A}'\mathcal{B}'}(\hat{\rho'}\hat{C}') = \langle \hat{C}'\rangle_{\rho'}$, where $\hat{\rho}' = Tr_{\mathcal{AB}\setminus\mathcal{A}'\mathcal{B}'}{\hat{\rho}}$.
What is the advantage of considering $\langle\hat{C}'\rangle_{\rho'}$? 
If we wanted to evaluate $\langle\hat{C}\rangle_\rho$, one would need to define a base of $\mathcal{H}_\mathrm{initial}$ (allowing one to represent $\hat{\rho}$ and $\hat{C}$). This consists in imposing all boundary conditions corresponding to boundary points of $\mathcal{A}$ and $\mathcal{B}$, and fixing the physical Gauss's law sector for internal points of $\mathcal{AB}$. In contrast, evaluating $\langle\hat{C}'\rangle_{\rho'}$ requires only a basis for $\mathcal{H}_\mathrm{final}$, which is constituted by less degrees of freedom ($\mathcal{A}'\subseteq \mathcal{A}$, $\mathcal{B}' \subseteq \mathcal{B}$), and with fewer constraints.

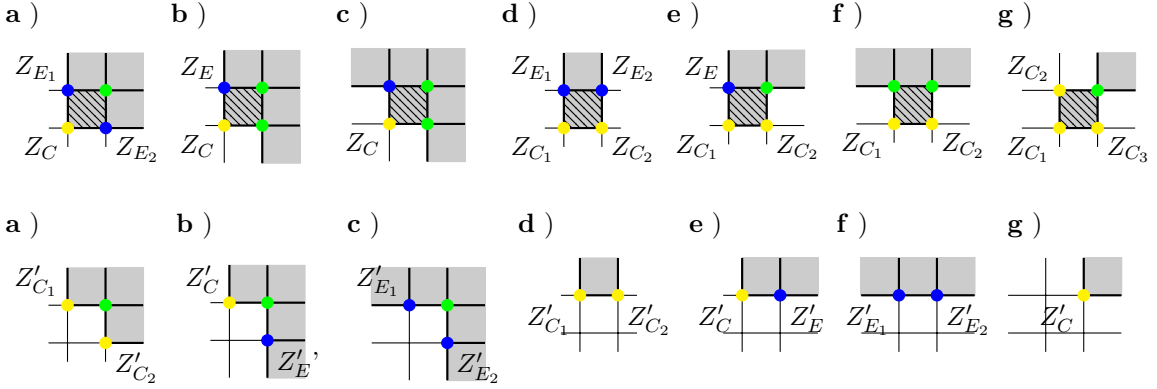
\begin{figure}[h]
    \centering
    \begin{subfigure}[t]{.12\textwidth}
    \caption{)}
    \begin{tikzpicture}[scale=.5]
        \draw[fill, color=gray!40] (0,0) rectangle (1.99,1.99);
        \draw[fill, pattern= north west lines] (0,0) rectangle (1,1);
        \draw[thin] (-.5,-.5) grid (1.99,1.99);
        \draw[thick] (0,0) grid (1.99,1.99);
        
        \def \r{.15}
        \begin{scope}[every node/.style={color=black,scale=.9}]
            \draw[fill,color=yellow] (0,0) circle (\r) node[anchor=north east]{$Z_{C}$};
            \draw[fill,color=green] (1,1) circle (\r); 
            \draw[fill,color=blue] (0,1) circle (\r) node[anchor= south east]{$Z_{E_1}$};
            \draw[fill,color=blue] (1,0) circle (\r) node[anchor= north west]{$Z_{E_2}$};
        \end{scope}
    \end{tikzpicture}
\end{subfigure}\hfill
\begin{subfigure}[t]{.12\textwidth}
    \caption{)}
    \begin{tikzpicture}[scale=.5]
        \draw[fill, color=gray!40] (0,0) rectangle (1.99,1.99);
        \draw[fill, color=gray!40] (1,0) rectangle (2,-1);
        
        \draw[fill, pattern= north west lines] (0,0) rectangle (1,1);
        \draw[thin] (-.5,-.99) grid (1.99,1.99);
        \draw[thick] (1,0) -- (1,-1);
        
        \draw[thick] (0,0) grid (1.99,1.99);
        
        \def \r{.15}
        \begin{scope}[every node/.style={color=black,scale=.9}]
            \draw[fill,color=yellow] (0,0) circle (\r) node[anchor=north east]{$Z_{C}$};
            \draw[fill,color=green] (1,1) circle (\r); 
            \draw[fill,color=blue] (0,1) circle (\r) node[anchor= south east]{$Z_{E}$};
            \draw[fill,color=green] (1,0) circle (\r);
        \end{scope}
    \end{tikzpicture}
\end{subfigure}\hfill
\begin{subfigure}[t]{.12\textwidth}
    \caption{)}
    \begin{tikzpicture}[scale=.5]
        \draw[fill, color=gray!40] (0,0) rectangle (1.99,1.99);
        \draw[fill, color=gray!40] (1,0) rectangle (2,-1);
        \draw[fill, color=gray!40] (0,1) rectangle (-1,2);
        \draw[fill, pattern= north west lines] (0,0) rectangle (1,1);
        \draw[thin] (-.99,-.99) grid (1.99,1.99);
        \draw[thick] (1,0) -- (1,-1);
        \draw[thick] (0,1) -- (-1,1);
        \draw[thick] (0,0) grid (1.99,1.99);
        
        \def \r{.15}
        \begin{scope}[every node/.style={color=black,scale=.9}]
            \draw[fill,color=yellow] (0,0) circle (\r) node[anchor=north east]{$Z_{C}$};
            \draw[fill,color=green] (1,1) circle (\r); 
            \draw[fill,color=blue] (0,1) circle (\r);
            \draw[fill,color=green] (1,0) circle (\r);
        \end{scope}
    \end{tikzpicture}
\end{subfigure}\hfill
\begin{subfigure}[t]{.12\textwidth}
    \caption{)}
    \begin{tikzpicture}[scale=.5]
        \draw[fill, color=gray!40] (0,0) rectangle (1,1.99);
        \draw[fill, pattern= north west lines] (0,0) rectangle (1,1);
        \draw[thin] (-.5,-.5) grid (1.5,1.99);
        \draw[thick] (0,0) grid (1,1.99);
        
        \def \r{.15}
        \begin{scope}[every node/.style={color=black,scale=.9}]
            \draw[fill,color=yellow] (0,0) circle (\r) node[anchor=north east]{$Z_{C_1}$};
            \draw[fill,color=blue] (1,1) circle (\r) node[anchor=south west]{$Z_{E_2}$}; 
            \draw[fill,color=blue] (0,1) circle (\r) node[anchor= south east]{$Z_{E_1}$};
            \draw[fill,color=yellow] (1,0) circle (\r) node[anchor= north west]{$Z_{C_2}$};
        \end{scope}
    \end{tikzpicture}
\end{subfigure}\hfill
\begin{subfigure}[t]{.12\textwidth}
    \caption{)}
    \begin{tikzpicture}[scale=.5]
        \draw[fill, color=gray!40] (0,0) rectangle (1,1.99);
        \draw[fill, color=gray!40] (1,1) rectangle (2,2);
        \draw[fill, pattern= north west lines] (0,0) rectangle (1,1);
        \draw[thin] (-.5,-.5) grid (1.99,1.99);
        \draw[thick] (0,0) grid (1,1.99);
        \draw[thick] (1,1)--(2,1);
        
        \def \r{.15}
        \begin{scope}[every node/.style={color=black,scale=.9}]
            \draw[fill,color=yellow] (0,0) circle (\r) node[anchor=north east]{$Z_{C_1}$};
            \draw[fill,color=green] (1,1) circle (\r); 
            \draw[fill,color=blue] (0,1) circle (\r) node[anchor= south east]{$Z_{E}$};
            \draw[fill,color=yellow] (1,0) circle (\r) node[anchor= north west]{$Z_{C_2}$};
        \end{scope}
    \end{tikzpicture}
\end{subfigure}\hfill
\begin{subfigure}[t]{.12\textwidth}  
    \caption{)}
    \begin{tikzpicture}[scale=.5]
        \draw[fill, color=gray!40] (0,0) rectangle (1,1.99);
        \draw[fill, color=gray!40] (1,1) rectangle (2,2);
        \draw[fill, color=gray!40] (0,1) rectangle (-1,2);
        \draw[fill, pattern= north west lines] (0,0) rectangle (1,1);
        \draw[thin] (-.99,-.5) grid (1.99,1.99);
        \draw[thick] (0,0) grid (1,1.99);
        \draw[thick] (1,1)--(2,1);
        \draw[thick] (0,1)--(-1,1);
        
        \def \r{.15}
        \begin{scope}[every node/.style={color=black,scale=.9}]
            \draw[fill,color=yellow] (0,0) circle (\r) node[anchor=north east]{$Z_{C_1}$};
            \draw[fill,color=green] (1,1) circle (\r); 
            \draw[fill,color=green] (0,1) circle (\r);
            \draw[fill,color=yellow] (1,0) circle (\r) node[anchor= north west]{$Z_{C_2}$};
        \end{scope}
    \end{tikzpicture}
\end{subfigure}\hfill
\begin{subfigure}[t]{.12\textwidth}
    \caption{)}
    \begin{tikzpicture}[scale=.5]
        \draw[fill, color=gray!40] (1,1) rectangle (2,2);
        \draw[fill, color=gray!40] (0,0) rectangle (1,1);
        \draw[fill, pattern= north west lines] (0,0) rectangle (1,1);
        \draw[thin] (-.99,-.5) grid (1.99,1.99);
        \draw[thick] (0,0) grid (1,1);
        \draw[thick] (1,1)--(2,1);
        \draw[thick] (1,1)--(1,2);
        
        \def \r{.15}
        \begin{scope}[every node/.style={color=black,scale=.9}]
            \draw[fill,color=yellow] (0,0) circle (\r) node[anchor=north east]{$Z_{C_1}$};
            \draw[fill,color=green] (1,1) circle (\r); 
            \draw[fill,color=yellow] (0,1) circle (\r) node[anchor= south east]{$Z_{C_2}$};
            \draw[fill,color=yellow] (1,0) circle (\r) node[anchor= north west]{$Z_{C_3}$};
        \end{scope}
    \end{tikzpicture}
    \label{fg:B:before}
\end{subfigure}
    \setcounter{subfigure}{0}
    
    \begin{subfigure}[t]{.12\textwidth}
    \caption{)}
    \begin{tikzpicture}[scale=.5]
        \draw[fill, color=gray!40] (0,1) rectangle (1.99,1.99);
        \draw[fill, color=gray!40] (1,0) rectangle (2,1.99);
        
        \draw[thin] (-.5,-.5) grid (1.99,1.99);
        \draw[thick] (0,1) grid (1.99,1.99);
        \draw[thick] (1,0) -- (1,1);
        \draw[thick] (1,0) -- (2,0);
        
        \def \r{.15}
        \begin{scope}[every node/.style={color=black,scale=.9}]
            \draw[fill,color=green] (1,1) circle (\r); 
            \draw[fill,color=yellow] (0,1) circle (\r) node[anchor= south east]{$Z'_{C_1}$};
            \draw[fill,color=yellow] (1,0) circle (\r) node[anchor= north west]{$Z'_{C_2}$};
        \end{scope}
    \end{tikzpicture}
\end{subfigure}
\hfill
\begin{subfigure}[t]{.12\textwidth}    
    \caption{)}
    \begin{tikzpicture}[scale=.5]
        \draw[fill, color=gray!40] (1,-1) rectangle (1.99,1.99);
        \draw[fill, color=gray!40] (1,1) rectangle (0,1.99);

        \draw[thin] (-.5,-.99) grid (1.99,1.99);
        \draw[thick] (1,-.99) grid (1.99,1.99);
        \draw[thick] (0,1) -- (0,2);
        \draw[thick] (0,1) -- (1,1);
        
        \def \r{.15}
        \begin{scope}[every node/.style={color=black,scale=.9}]
            \draw[fill,color=green] (1,1) circle (\r); 
            \draw[fill,color=yellow] (0,1) circle (\r) node[anchor= south east]{$Z'_{C}$};
            \draw[fill,color=blue] (1,0) circle (\r) node[anchor=north west]{$Z'_E$'};
        \end{scope}
    \end{tikzpicture}
\end{subfigure}
\hfill
\begin{subfigure}[t]{.12\textwidth}
    \caption{)}
    \begin{tikzpicture}[scale=.5]
        \draw[fill, color=gray!40] (-1,1) rectangle (2,2);
        \draw[fill, color=gray!40] (1,-1) rectangle (2,1);
        
        \draw[thin] (-.99,-.99) grid (1.99,1.99);
        \draw[thick] (-.99,1) grid (1.99,1.99);
        \draw[thick] (1,-.99) grid (1.99,.99);
        
        \def \r{.15}
        \begin{scope}[every node/.style={color=black,scale=.9}]
            \draw[fill,color=green] (1,1) circle (\r); 
            \draw[fill,color=blue] (0,1) circle (\r) node[anchor=south east]{$Z'_{E_1}$};
            \draw[fill,color=blue] (1,0) circle (\r) node[anchor=north west]{$Z'_{E_2}$};
        \end{scope}
    \end{tikzpicture}
\end{subfigure}
\hfill
\begin{subfigure}[t]{.12\textwidth}
    \caption{)}
    \begin{tikzpicture}[scale=.5]
    \draw[fill, color=gray!40] (0,1) rectangle (1,1.99);
    
    \draw[thin] (-.5,-.5) grid (1.5,1.99);
    \draw[thick] (0,1) grid (1,1.99);
    
    \def \r{.15}
    \begin{scope}[every node/.style={color=black,scale=.9}]
        \draw[fill,color=yellow] (1,1) circle (\r) node[anchor=north west]{$Z'_{C_2}$}; 
        \draw[fill,color=yellow] (0,1) circle (\r) node[anchor= north east]{$Z'_{C_1}$};
    \end{scope}
    \end{tikzpicture}
\end{subfigure}
\hfill
\begin{subfigure}[t]{.12\textwidth}
    \caption{)}
    \begin{tikzpicture}[scale=.5]
        \draw[fill, color=gray!40] (0,1) rectangle (2,2);
        
        \draw[thin] (-.5,-.5) grid (1.99,1.99);
        \draw[thick] (0,1) grid (1.99,1.99);
        
        \def \r{.15}
        \begin{scope}[every node/.style={color=black,scale=.9}]
            \draw[fill,color=blue] (1,1) circle (\r) node[anchor=north west]{$Z'_E$}; 
            \draw[fill,color=yellow] (0,1) circle (\r) node[anchor= north east]{$Z'_{C}$};
        \end{scope}
    \end{tikzpicture}
\end{subfigure}
\begin{subfigure}[t]{.12\textwidth}
    \caption{)}
    \begin{tikzpicture}[scale=.5]
        \draw[fill, color=gray!40] (-1,1) rectangle (2,2);
        
        \draw[thin] (-.99,-.5) grid (1.99,1.99);
        \draw[thick] (-.99,1) grid (1.99,1.99);
        
        \def \r{.15}
        \begin{scope}[every node/.style={color=black,scale=.9}]
            \draw[fill,color=blue] (1,1) circle (\r) node[anchor= north west]{$Z'_{E_2}$}; 
            \draw[fill,color=blue] (0,1) circle (\r) node[anchor=north east]{$Z'_{E_1}$};
            
        \end{scope}
    \end{tikzpicture}
\end{subfigure}
\hfill    
\begin{subfigure}[t]{.12\textwidth}
    \caption{)}
    \begin{tikzpicture}[scale=.5]
        \draw[fill, color=gray!40] (1,1) rectangle (2,2);
        
        \draw[thin] (-.99,-.5) grid (1.99,1.99);
        \draw[thick] (1,1) grid (1.99,1.99);
        
        \def \r{.15}
        \begin{scope}[every node/.style={color=black,scale=.9}]
            \draw[fill,color=yellow] (1,1) circle (\r) node[anchor=north east]{$Z'_C$};
        \end{scope}
    \end{tikzpicture}
    \label{fg:B:after}
\end{subfigure}
    
    \caption{Reduction scheme applied to different cases. Top: Different initial cases: corner boundary points (yellow), edge boundary points (blue), internal points (green). Edges belonging to $\mathcal{X}$ are thickened. Bottom: Final cases obtained reducing cases in figure~\ref{fg:B:before}. The edges belonging to $\mathcal{X}'$ (gray shaded region) are thickened.}
\end{figure}

In the following, we show the procedure that allows us, given a region $\mathcal{X}$ with border conditions encoded in $Z_\mathcal{X}$, to recover a subregion $\mathcal{X}'$, containing less unflippable plaquettes, and the associated border conditions set by $Z'_{\mathcal{X}'}$.
We can distinguish 3 kinds of vertices in $\mathcal{X}$: $C$ is a corner boundary point, where $Z_C$ takes values $0, \pm 1$; $E$ is an edge boundary point, where $Z_E$ takes values $\pm \nicefrac{1}{2}$; and $I$ is an internal point if it does not belong to $\partial_0\mathcal{X}$.
Corner boundary points that have values $\pm 1$ are part of an unflippable plaquette, which does not contribute to the witness defined in equation~\eqref{eq:gauge:witness} and can thus be removed. In figure~\ref{fg:B:before} we represent all possible initial situations. Unflippable plaquettes can be removed from the considerations by applying the transformation rules to the superselection sectors summarized in table~\ref{tab:reduction_scheme}, where we introduce $\theta_1(x_1,x_2) = \frac{\delta_{x_1,0}}{2} (x_2 - x_1) + \frac{x_1}{2}$ and 
$\theta_2(x_1,x_2) = \frac{\delta_{x_2,0}}{2}(x_1 - x_2) + \frac{x_2}{2}$. We end up with situations sketched in figure~\ref{fg:B:after}, which can be considerably simpler, depending on the number of unflippable plaquettes that can be removed.

\begin{table}[h]
    \centering
    \small
    \begin{tabular}{| m{2.5cm}  m{2.5cm}  m{3cm} |}
        \hline
         & Quantity & Expression \\\hline
        \multirow{2}{*}{Case \textbf{a}} & $Z'_{C_1}$ & $ Z_{E_1} + \nicefrac{Z_C}{2}$\\
            & $Z'_{C_2}$ & $Z_{E_2}+\nicefrac{Z_C}{2}$ \\[.3cm]
        \multirow{2}{*}{Case \textbf{b}} 
            & $Z'_C$ & $Z_E + \nicefrac{Z_C}{2}$\\
            & $Z'_E$ & $\nicefrac{Z_C}{2}$ \\[.3cm]
        \multirow{1}{*}{Case \textbf{c}} 
            & $Z'_{E_1} = Z'_{E_2}$ & $\nicefrac{Z_C}{2}$ \\[.3cm]
        \multirow{2}{*}{Case \textbf{d}}
            & $Z'_{C_1}$ & $Z_{E_1} + \theta_1(Z_{C_1},Z_{C_2})$\\
            & $Z'_{c_2}$ & $ Z_{E_2} + \theta_2(Z_{C_1},Z_{C_2})$\\[.3cm]
    \multirow{2}{*}{Case \textbf{e}} 
        & $Z'_{C}$ & $Z_E + \theta_1(Z_{c,1},Z_{C_2})$ \\
        & $Z'_{E}$ & $\theta_2(Z_{C_1},Z_{C_2})$\\[.3cm]
    \multirow{2}{*}{Case \textbf{f}}
        & $Z'_{E_1}$ & $\theta_1(Z_{C_1},Z_{C_2})$\\
        & $Z'_{E_2}$ & $\theta_2(Z_{C_1},Z_{C_2})$\\[.3cm]
    \multirow{1}{*}{Case \textbf{g}} & $Z'_C$ & $Z_{C_1}+Z_{C_2}+Z_{C_3}$\\
    \hline
    \end{tabular}
    \caption[justification=justified]{Relations between $Z_\mathcal{X}$ and $Z'_{\mathcal{X}'}$ for the different cases.}
    \label{tab:reduction_scheme}
\end{table}

\section{Optimization method}
\label{sec:appendix_C}
In this section, we introduce a numerical technique to evaluate bounds of $\hat{C}$ when evaluated on a separable state, given a fixed SS. 
For a given basis choice, we denote with $\mathcal{M}^\mathcal{A}$ the set of matrices  $[{\mathcal{U}}^\mathcal{A}_{p_1},\dots,{\mathcal{U}}^\mathcal{A}_{p_n},{\mathcal{P}}^\mathcal{A}_{\ell_1},\dots,{\mathcal{P}}^\mathcal{A}_{\ell_m}]$ that represent operators of equation~\eqref{eq:matter:witness} acting on \(\lat{A}\).
Similarly, we define $\mathcal{M}^\mathcal{B} = [{\mathcal{U}}^\mathcal{B}_{p_1},\dots,{\mathcal{U}}^\mathcal{B}_{p_n},{\mathcal{P}}^\mathcal{B}_{\ell_1},\dots,{\mathcal{P}}^\mathcal{B}_{\ell_m}]$.
In the situation under consideration, the matrices in $\mathcal{M}^\mathcal{A}$ and in $\mathcal{M}^\mathcal{B}$ have real entries.
For the chosen basis, a separable state between $\mathcal{A}$ and $\mathcal{B}$ is represented through two complex vectors ${\alpha} = \{a_i e^{i\phi^\mathcal{A}_i}\}_i$ and ${\beta} = \{b_ie^{i\phi^\mathcal{B}_i}\}_i$.
Expanding $\braket{\hat{C}}_\mathrm{sep}\equiv \bra{\Psi_\mathrm{sep}}\hat{C}\ket{\Psi_\mathrm{sep}}$, we find
\begin{align}
    \braket{\hat{C}}_\mathrm{sep}=&
     4\sum_{n}(\sum_{i,j} a_i \mathcal{M}^{\mathcal{A},n}_{i,j} a_j \cos(\phi^\mathcal{A}_j-\phi^\mathcal{A}_i)) 
     \times (\sum_{l,m} b_l  \mathcal{M}^{\mathcal{B},n}_{l,m} b_m \cos(\phi^\mathcal{B}_l-\phi^\mathcal{B}_m)) \,. 
\end{align}
Using the triangular inequality and the fact that the $\cos(\cdot)$ function takes values in $(-1,+1)$, we conclude
\begin{align}
    \lvert\langle\hat{C}\rangle_{\mathrm{sep}}\rvert \leq \sum_n (\sum_{i,j} a_i{\mathcal{M}^{\mathcal{A},n}_{i,j}a_j} )
    (\sum_{l,m}b_l {M}^{\mathcal{B},n}_{l,m} b_m)\,
\end{align}
with $a_i,b_i \in (0,1)$ and the normalization constraints $\sum_i a_i^2 =\sum_j b_j^j = 1$.
Using the method of Lagrange multipliers, one finds the system of equations
\begin{subequations}
\label{eq:optimizationsystem}
\begin{align}
        \mathcal{M}^\mathcal{A}(b)\cdot a = &\lambda_\mathcal{A}a\\
        \mathcal{M}^\mathcal{B}(a)\cdot b = &\lambda_\mathcal{B}(b)\\
        a^Ta = &1\\
        b^Tb = &1\,,
\end{align}
\end{subequations}
where we defined $\mathcal{M}^\mathcal{A}(b)=\sum_n (b^T\mathcal{M}^{\mathcal{B},n}b) \mathcal{M}^{\mathcal{A},n}$ and equivalently $\mathcal{M}^\mathcal{B}(a)$.  
It is possible to verify that $a^T \mathcal{M}^\mathcal{A}(b)\cdot a = \lambda_\mathcal{A} = f(\bm{a},\bm{b})=b^T\mathcal{M}^\mathcal{B}(a)\cdot b = \lambda_\mathcal{B}$. 

To solve the equation system~\eqref{eq:optimizationsystem} for large regions where analytic computations become unfeasible, we implement an iterative method.
We initialize $a_0$ to a random normalized vector, evaluate $\mathcal{M}^\mathcal{B}(a_0) = \sum_n (a_0^T \mathcal{M}^{\mathcal{A},n}a_0) \mathcal{M}^{\mathcal{B},n}$ and solve the eigenvalue problem $\mathcal{M}^\mathcal{B}(a_0) b_0 = \lambda_\mathcal{B}b_0$.
Since we are looking for a maximum of $f(\cdot,\cdot)$, we choose the eigenstate with maximal eigenvalue $\lambda_\mathcal{B}$.
Then, we use $b_0$ to evaluate $\mathcal{M}^{\mathcal{A}}(b_0) = \sum_n (b_0^T\mathcal{M}^{\mathcal{B},n}b_0)\mathcal{M}^{\mathcal{A},n}$ and solve the eigenvalue problem $\mathcal{M}^\mathcal{A}(b_0) a_1 = \lambda_\mathcal{A}a_1$, selecting again the eigenstate with maximal eigenvalue $\lambda_\mathcal{A}$.
This process is iterated until convergence. In the considered scenarios, we find that only few trials converge to a local maximum rather than the global maximum of $f(\cdot,\cdot)$. 
Since the system studied in \ref{sec:matter:witness} is small, we optimized $f(\cdot,\cdot)$ analytically.
For more complex systems, this numerical method can be successfully used to derive bounds.

\end{document}